\documentclass[preprint, astrosymb]{aastex631}
\usepackage{CJK}
\usepackage{amsmath}
\hypersetup{linkcolor=blue,citecolor=blue,filecolor=blue,urlcolor=blue}

\begin{document}
\begin{CJK*}{UTF8}{gbsn}
\shorttitle{Multiscale velocities in NGC 6334}
\shortauthors{Liu et al.}
\title{Deviation from a Continuous and Universal Turbulence Cascade in NGC 6334 due to Massive Star Formation Activity}

\correspondingauthor{Junhao Liu}
\email{liujunhao42@outlook.com; j.liu@eaobservatory.org}

\author[0000-0002-4774-2998]{Junhao Liu (刘峻豪)}
\affiliation{East Asian Observatory, 660 N. A`oh\={o}k\={u} Place, University Park, Hilo, HI 96720, USA}

\author[0000-0003-2384-6589]{Qizhou Zhang}
\affiliation{Center for Astrophysics $\vert$ Harvard \& Smithsonian, 60 Garden Street, Cambridge, MA 02138, USA}

\author[0000-0003-2300-2626]{Hauyu Baobab Liu}
\affiliation{Physics Department, National Sun Yat-Sen University, No. 70, Lien-Hai Road, Kaohsiung City 80424, Taiwan, Republic of China}
\affiliation{Institute of Astronomy and Astrophysics, Academia Sinica, 11F of Astronomy-Mathematics Building, AS/NTU No.1, Sec. 4, Roosevelt Rd, Taipei 10617, Taiwan, Republic of China} 

\author[0000-0002-5093-5088]{Keping Qiu}
\affiliation{School of Astronomy and Space Science, Nanjing University, 163 Xianlin Avenue, Nanjing 210023, Jiangsu, People's Republic of China}
\affiliation{Key Laboratory of Modern Astronomy and Astrophysics (Nanjing University), Ministry of Education, Nanjing 210023, Jiangsu, People's Republic of China}

\author[0000-0003-1275-5251]{Shanghuo Li}
\affiliation{Max Planck Institute for Astronomy, Konigstuhl 17, D-69117 Heidelberg, Germany}

\author[0000-0002-7402-6487]{Zhi-Yun Li}
\affiliation{Astronomy Department, University of Virginia, Charlottesville, VA 22904-4325, USA}

\author[0000-0002-3412-4306]{Paul T. P. Ho}
\affiliation{East Asian Observatory, 660 N. A`oh\={o}k\={u} Place, University Park, Hilo, HI 96720, USA}
\affiliation{Institute of Astronomy and Astrophysics, Academia Sinica, 11F of Astronomy-Mathematics Building, AS/NTU No.1, Sec. 4, Roosevelt Rd, Taipei 10617, Taiwan, Republic of China}

\author[0000-0002-3829-5591]{Josep Miquel Girart}
\affil{Institut de Ci\`{e}ncies de l'Espai (ICE, CSIC), Can Magrans s/n, E-08193 Cerdanyola del Vall\`{e}s, Catalonia, Spain}
\affil{Institut d'Estudis Espacials de de Catalunya (IEEC), E-08034 Barcelona, Catalonia, Spain}

\author[0000-0001-8516-2532]{Tao-Chung Ching}
\affiliation{National Radio Astronomy Observatory, P.O. Box O, Socorro, NM 87801, USA}

\author[0000-0002-9774-1846]{Huei-Ru Vivien Chen}
\affiliation{Institute of Astronomy and Astrophysics, Academia Sinica, 11F of Astronomy-Mathematics Building, AS/NTU No.1, Sec. 4, Roosevelt Rd, Taipei 10617, Taiwan, Republic of China}

\author[0000-0001-5522-486X]{Shih-Ping Lai}
\affiliation{Institute of Astronomy and Department of Physics, National Tsing Hua University, Hsinchu 30013, Taiwan, Republic of China}
\affiliation{Institute of Astronomy and Astrophysics, Academia Sinica, 11F of Astronomy-Mathematics Building, AS/NTU No.1, Sec. 4, Roosevelt Rd, Taipei 10617, Taiwan, Republic of China}

\author[0000-0002-1407-7944]{Ramprasad Rao}
\affiliation{Institute of Astronomy and Astrophysics, Academia Sinica, 11F of Astronomy-Mathematics Building, AS/NTU No.1, Sec. 4, Roosevelt Rd, Taipei 10617, Taiwan, Republic of China}
\affiliation{Center for Astrophysics $\vert$ Harvard \& Smithsonian, 60 Garden Street, Cambridge, MA 02138, USA}

\author[0000-0002-0675-276X]{Ya-wen Tang}
\affiliation{Institute of Astronomy and Astrophysics, Academia Sinica, 11F of Astronomy-Mathematics Building, AS/NTU No.1, Sec. 4, Roosevelt Rd, Taipei 10617, Taiwan, Republic of China}

\begin{abstract}

We use molecular line data from ALMA, SMA, JCMT, and NANTEN2 to study the multi-scale ($\sim$15-0.005 pc) velocity statistics in the massive star formation region NGC 6334. We find that the non-thermal motions revealed by the velocity dispersion function (VDF) stay supersonic over scales of several orders of magnitudes. The multi-scale non-thermal motions revealed by different instruments do not follow the same continuous power-law, which is because the massive star formation activities near central young stellar objects have increased the non-thermal motions in small-scale and high-density regions. The magnitudes of VDFs vary in different gas materials at the same scale, where the infrared dark clump N6334S in an early evolutionary stage shows a lower level of non-thermal motions than other more evolved clumps due to its more quiescent star formation activity. We find possible signs of small-scale-driven (e.g., by gravitational accretion or outflows) supersonic turbulence in clump N6334IV with a three-point VDF analysis. Our results clearly show that the scaling relation of velocity fields in NGC 6334 deviates from a continuous and universal turbulence cascade due to massive star formation activities. 

\end{abstract}



\section{Introduction} \label{sec:intro}
Turbulence plays an important role in the star formation process within molecular clouds\footnote{The nomenclature of cloud ($\sim$10 pc), clump ($\sim$1 pc), core ($\sim$0.1 pc), and condensation ($\sim$0.01 pc) scales in this paper follows \citet{2009ApJ...696..268Z}.} by providing support against gravitational collapse and regulating the hierarchical fragmentation of cloud substructures \citep{2004RvMP...76..125M}. One important property of turbulence is the relation of velocity statistics at different spatial scales. Since the first attempt by \citet{1981MNRAS.194..809L}, the scaling of the velocity structure has been extensively studied with the linewidth-size relation from molecular line observations. At $>$0.1 pc scales, observational studies usually found a power-law scaling for the linewidth-size relation \citep[e.g.,][]{1983ApJ...270..105M, 1987ApJ...319..730S, 1995ApJ...446..665C}, which was interpreted as a natural consequence of turbulence cascade. At $<$0.1 pc scales, the linewidth-size relation could transit to ``coherent'' (i.e., constant linewidth) for low-mass dense cores due to the dominance of thermal motions over non-thermal motions \citep{1998ApJ...504..223G}, but the relation in high-mass star formation regions is less clear due to rarer observations. 

Other than the linewidth-size relation, the dispersion function (autocorrelation function or structure function) of velocity centroids can also be used to study the scaling relation of velocity fields in molecular clouds \citep[e.g., ][]{1985ApJ...295..479D, 1994ApJ...429..645M}. Compared to the linewidth, the velocity centroid is advantageous in revealing  the non-thermal motions because it is not sensitive to thermal broadening \citep{2022ApJ...935...77L}. The velocity centroid also does not suffer from the effect of line-of-sight signal integration that could broaden the observed linewidth \citep{2016ApJ...821...21C}. However, the dispersion function of velocity centroids is relatively less applied to observations due to its complexity. 

NGC 6334 is a nearby \citep[$\sim$1.3 kpc,][]{2014ApJ...784..114C, 2014A&A...566A..17W} massive star formation region. Six massive clumps (N6334I-V and N6334I(N))  with active massive star formation activities were identified in the predominant 10 pc-long NGC 6334 filament located at the center of the NGC 6334 region with far-infrared/sub-mm/mm observations \citep[e.g., ][]{1978ApJ...226L.149C, 1979ApJ...232L.183M, 1982ApJ...259L..29G}. In addition, a massive infrared dark clump N6334S is located near the southwestern end of the NGC 6334 region. In this paper, we present a multi-scale study of the velocity statistics in 5 massive clumps (N6334I(N), I, IV, V, and S) as well as their parental structures in NGC 6334 using velocity centroid structure functions of different molecular line tracers from different instruments.

\section{Observation} \label{sec:observation}
A particular line tracer is insensitive to gas below its critical density and is only sensitive to densities at most a factor of 2 orders of magnitude above its critical density \citep{1998ApJ...504..223G}, so we use different line tracers to trace gas materials at different densities. The information on the molecular line data is summarized in Table \ref{tab:line}. The critical density ($n_{\mathrm{c}}$) of each line is estimated with $n_{\mathrm{c}} = A_{ul}/q_{ul}(T)$, where $A_{ul}$, $q_{ul}(T)$, and $T$ are the Einstein coefficient, collision rate coefficient, and gas temperature, respectively. Values of $A_{ul}$ and $q_{ul}$ are adopted from the CDMS \citep{2001AA...370L..49M} and LAMDA \citep{2005AA...432..369S} databases. No information on the collision rate coefficient exists for $^{13}$CS and H$^{13}$CO$^+$ in LAMDA, so we adopt the values for CS and HCO$^+$ instead. Detailed observational parameters for each line can be found in the subsections below. 

\begin{deluxetable}{cccccccc}[t!]
\tablecaption{Summary of molecular line data \label{tab:line}}
\tablecolumns{8}
\tablewidth{0pt}
\tablehead{
\colhead{Line} &
\colhead{Frequency} &
\colhead{$E_{\mathrm{u}}/k$ \tablenotemark{a}} & 
\colhead{$n_{\mathrm{c}}$ \tablenotemark{b}} &
\colhead{Instrument} & 
\colhead{$l_{\mathrm{beam}}$  \tablenotemark{c}} & 
\colhead{$l_{\mathrm{MRS}}$  \tablenotemark{d}} & 
\colhead{Targets} \\
\colhead{} &  \colhead{(GHz)} & \colhead{(K)} &  \colhead{(cm$^{-3}$)}  & \colhead{} & \colhead{($\arcsec$)}  & \colhead{($\arcsec$)} & \colhead{}
}
\startdata
OCS (19-18) & 231.0610 & 110.9 & 4.8 $\times$ 10$^5$ & ALMA & 0.7 & 13 & I(N), I, IV, V\\ 
$^{13}$CS (5-4) & 231.2207 & 33.3 & 5.4 $\times$ 10$^6$ & ALMA &  0.7 & 13 & I(N), I, IV, V\\ 
H$^{13}$CO$^+$ (1-0) & 86.7543 & 4.2 & 1.8 $\times$ 10$^5$ & ALMA & 4.1 & 30 & S\\
NH$_2$D (1$_{1,1}$-1$_{0,1}$) & 85.9263 & 20.7 & 6.5 $\times$ 10$^4$ & ALMA & 4.1 & 30 & S\\ 
H$^{13}$CO$^+$ (4-3) & 346.9983 & 41.6 & 9.2 $\times$ 10$^6$ & SMA & 4 & 20 & I(N), I, IV, V\\
$^{13}$CO (3-2) & 330.5880 & 31.7 & 3.1 $\times$ 10$^4$ & JCMT & 14 & ... & I(N), I\\ 
$^{12}$CO (1-0) & 115.2712 & 5.5 & 2.2 $\times$ 10$^3$ & NANTEN2 & 180 & ... & Cloud\\ 
\enddata
\tablenotetext{}{Notes. The line information is from the CDMS \citep{2001AA...370L..49M} and LAMDA \citep{2005AA...432..369S} databases.}
\tablenotetext{a}{Upper energy level in units of K.}
\tablenotetext{b}{Critical density at specific temperatures (see Section \ref{sec:Ms}). For $^{13}$CS and H$^{13}$CO$^+$, we adopt the collision rate coefficient of the main isotope. For ALMA OCS (19-18) and $^{13}$CS (5-4) and SMA H$^{13}$CO$^+$ (4-3) lines, we adopt $T_\mathrm{gas}\sim$100 K in the calculation of $n_{\mathrm{c}}$. }
\tablenotetext{c}{Size of the beam for the single-dish or major size of the synthesized beam for the interferometer.}
\tablenotetext{d}{Maximum recoverable scale for interferometers.}

\end{deluxetable}

\subsection{ALMA Observation}
We carried out ALMA Band 6 dust continuum and molecular line observations toward four clumps (N6334I(N), I, IV, and V) in the massive region NGC 6334 in C43-1 and C43-4 configurations under the project 2017.1.00793.S (PI: Qizhou Zhang). Part of the data was reported in \citet{2023ApJ...945..160L}. We adopt the OCS (19-18) and $^{13}$CS (5-4) line data from the ALMA observations. The spatial resolution ($l_{\mathrm{beam}}$) of the combined (C43-1 and C43-4) images is $\sim0.7$\arcsec ($\sim$0.004 pc at a distance of 1.3 kpc). The maximum recoverable scale\footnote{https://almascience.eso.org/observing/observing-configuration-schedule/prior-cycle-observing-and-configuration-schedule} ($l_{\mathrm{MRS}}$) is $\sim 13 \arcsec$ ($\sim$0.08 pc). The RMS noises of the spectral line cubes with a velocity channel width of 0.16 km s$^{-1}$ are $\sim$3.8, 8.7, 3.0, and 5.2 mJy beam$^{-1}$ for N6334I(N), I, IV, and V, respectively. More details on the ALMA data can be found in \citet{2023ApJ...945..160L}. The high critical densities of OCS (19-18) and $^{13}$CS (5-4) make them suitable for studying the kinematics in the molecular dense cores/condensations in N6334I(N), I, IV, and V \citep[$n \sim 10^{7}$-$10^{9}$ cm$^{-3}$;][]{2020A&A...635A...2S, 2021ApJ...912..159P} except for the highest density fragments. The high upper energy level ($E_{\mathrm{u}}/k$) of OCS (19-18) makes it sensitive to the hot gas within those dense structures, while $^{13}$CS (5-4) could trace colder materials due to its relatively lower $E_{\mathrm{u}}/k$.

We adopt the H$^{13}$CO$^+$ (1-0) and NH$_2$D (1$_{1,1}$-1$_{0,1}$) line data (ID: 2016.1.00951.S, PI: Shaye Strom) from \citet{2020ApJ...896..110L} and \citet{2022ApJ...926..165L}. The spatial and spectral resolutions are $\sim$4.1\arcsec ($\sim$0.026 pc) and 0.21 km s$^{-1}$, respectively. The maximum recoverable scale is $\sim 30 \arcsec$ ($\sim$0.19 pc). The RMS noise level is 6 mJy beam$^{-1}$ per channel. The high critical densities of H$^{13}$CO$^+$ (1-0) and NH$_2$D (1$_{1,1}$-1$_{0,1}$) make them suitable for studying the kinematics in the molecular dense cores in N6334S \citep[$n \sim 10^{5}$-$10^{8}$ cm$^{-3}$;][]{2020ApJ...896..110L} except for the highest density fragments. The $E_{\mathrm{u}}/k$ of the two lines is not far from the temperature of N6334S \citep[$\sim$15 K;][]{2020ApJ...896..110L}. 

\subsection{SMA Data}
We adopt the H$^{13}$CO$^+$ (4-3) line data of N6334I(N), I, IV, and V from SMA observations that were reported in \citet{2014ApJ...792..116Z} and \citet{2021ApJ...912..159P}. The spatial resolution is $\sim4$\arcsec ($\sim$0.025 pc) for N6334V and $\sim2$\arcsec ($\sim$0.013 pc) for the other clumps. The maximum recoverable scale is $\sim 20 \arcsec$ ($\sim$0.13 pc) as reported by \citet{2014ApJ...792..116Z}. The spectral resolution is 0.7 km s$^{-1}$. The RMS noise of the 0.7 km s$^{-1}$ interval line cubes is $\sim$0.2 Jy beam$^{-1}$. The high critical density of H$^{13}$CO$^+$ (4-3) makes it suitable for studying the kinematics in molecular dense cores/condensations in N6334I(N), I, IV, and V \citep[$n \sim 10^{7}$-$10^{9}$ cm$^{-3}$;][]{2020A&A...635A...2S, 2021ApJ...912..159P}, but its relatively low $E_{\mathrm{u}}/k$ makes it less sensitive to hot gas materials. 

\subsection{JCMT Data}
We include in our analysis the $^{13}$CO (3-2) line cubes toward N6334I(N) and N6334I taken with the Heterodyne Array Receiver Program and Auto-Correlation Spectrometer and Imaging System \citep[HARP and ACSIS, ][]{2009MNRAS.399.1026B} from the JCMT data archive (program code: M11BN07). The $^{13}$CO (3-2) data has been previously reported in \citet{2023ApJ...945..160L}. The spatial and spectral resolutions of the $^{13}$CO (3-2) data are $\sim$14\arcsec ($\sim$0.09 pc) and 0.055 km s$^{-1}$, respectively. The map size is $2\arcmin \times 2\arcmin$ ($\sim$0.76 pc $\times$ 0.76 pc) for each field. The RMS noises of N6334I(N) and N6334I are 1.2 and 0.6 K per channel, respectively, in antenna radiation temperature ($T_{\mathrm{R}}^\ast$). The critical density of $^{13}$CO (3-2) is suitable for studying the kinematics in the northern hub (covering N6334I(N) and I) of the NGC 6334 cloud \citep[$n\sim$ several times of $10^5$ cm$^{-3}$;][]{2016A&A...592A..54A, 2020A&A...635A...2S, 2021A&A...647A..78A}.

\subsection{NANTEN2 Data}
We also include in our analysis the NANTEN2 $^{12}$CO (1-0) data from \citet{2018PASJ...70S..41F}. The spatial and spectral resolutions of the $^{12}$CO (1-0) cubes are $\sim$3$\arcmin$ ($\sim$1.1 pc) and 0.16 km s$^{-1}$, respectively. The typical RMS noise level is $\sim$1.2 K per channel. The appropriate critical density of $^{12}$CO (1-0) makes it suitable for studying the kinematics of the NGC 6334 cloud and its surrounding material in the same complex \citep[$n \sim 3.2 \times 10^3$ cm$^{-3}$;][]{1999ApJS..124..439K}.

\section{Results} \label{sec:results}

\subsection{Molecular lines and velocity centroids}\label{sec:line}

We use NANTEN2 $^{12}$CO (1-0), JCMT $^{13}$CO (3-2), SMA H$^{13}$CO$^+$ (4-3), and ALMA H$^{13}$CO$^+$ (1-0), NH$_2$D (1$_{1,1}$-1$_{0,1}$), OCS (19-18), and $^{13}$CS (5-4) line data to reveal the kinematics in NGC 6334 at different densities and scales. We estimate the velocity centroid $V_c(\boldsymbol{x})$ at position $\boldsymbol{x}$ with 
\begin{equation}
V_c(\boldsymbol{x}) = \frac{\Sigma_i^{N_{\mathrm{ch}}} I_i(\boldsymbol{x}) v_i \Delta v_{\mathrm{ch}}}{\Sigma_i^{N_{\mathrm{ch}}} I_i(\boldsymbol{x}) \Delta v_{\mathrm{ch}}},
\end{equation}
where $I_i(\boldsymbol{x})$, $v_i$, $\Delta v_{\mathrm{ch}}$, and $N_{\mathrm{ch}}$ are the line intensity, line-of-sight velocity, channel width, and number of integrated channels, respectively. The multi-scale intensity-weighted velocity centroid (i.e., moment 1) maps are shown in Appendix \ref{sec:linem1}. For the NANTEN2 and JCMT observations, we only consider the line emission from -12 to 4 km s$^{-1}$ which covers the main velocity component at these scales \citep{2018PASJ...70S..41F, 2022A&A...660A..56A}. For the SMA and ALMA observations, we only consider velocities within $\sim$5 km s$^{-1}$ \citep[$\sim$4 km s$^{-1}$ for N6334S to be consistent with][]{2020ApJ...896..110L} with respect to the local-standard-of-rest (LSR) velocity of the clump to avoid the contamination from outflows \citep{2023ApJ...945..160L}. The LSR velocities are $\sim$-3.5, -7.5, -3.5, and -6 km s$^{-1}$ for N6334I(N), I, IV, and V, respectively, rounded to the nearest 0.5 decimal \citep{2023ApJ...945..160L}. Ordered velocity gradients in NGC 6334 have been previously reported across different spatial scales \citep{2017ApJ...844...44J, 2018PASJ...70S..41F, 2022A&A...660A..56A, 2023ApJ...945..160L}.

\subsection{Velocity centroid dispersion functions}\label{sec:vdf} 
\subsubsection{Two-point velocity centroid dispersion functions}\label{sec:vdf2pt} 

\begin{figure*}[!htbp]
 \gridline{\fig{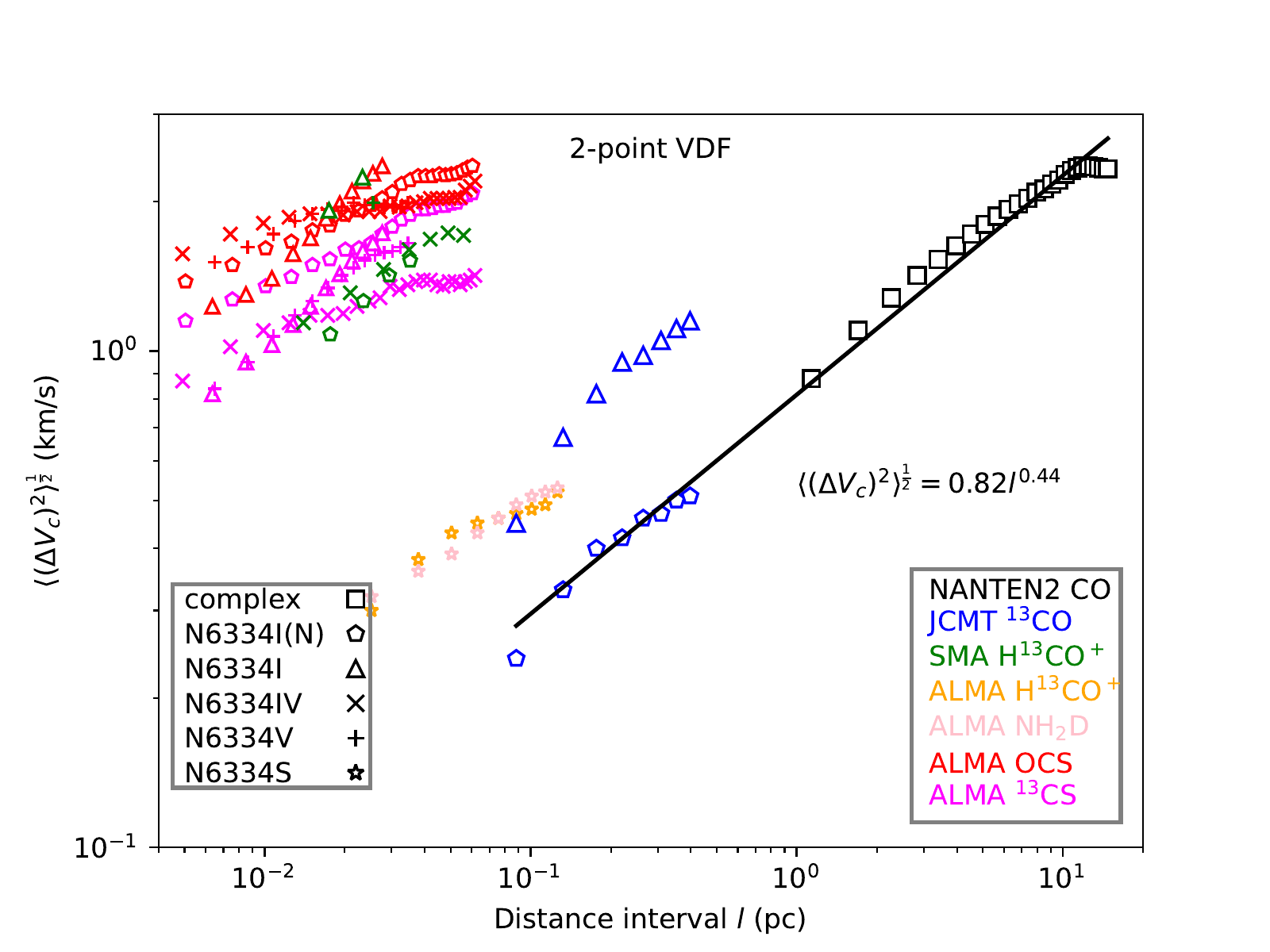}{0.48\textwidth}{}
 \fig{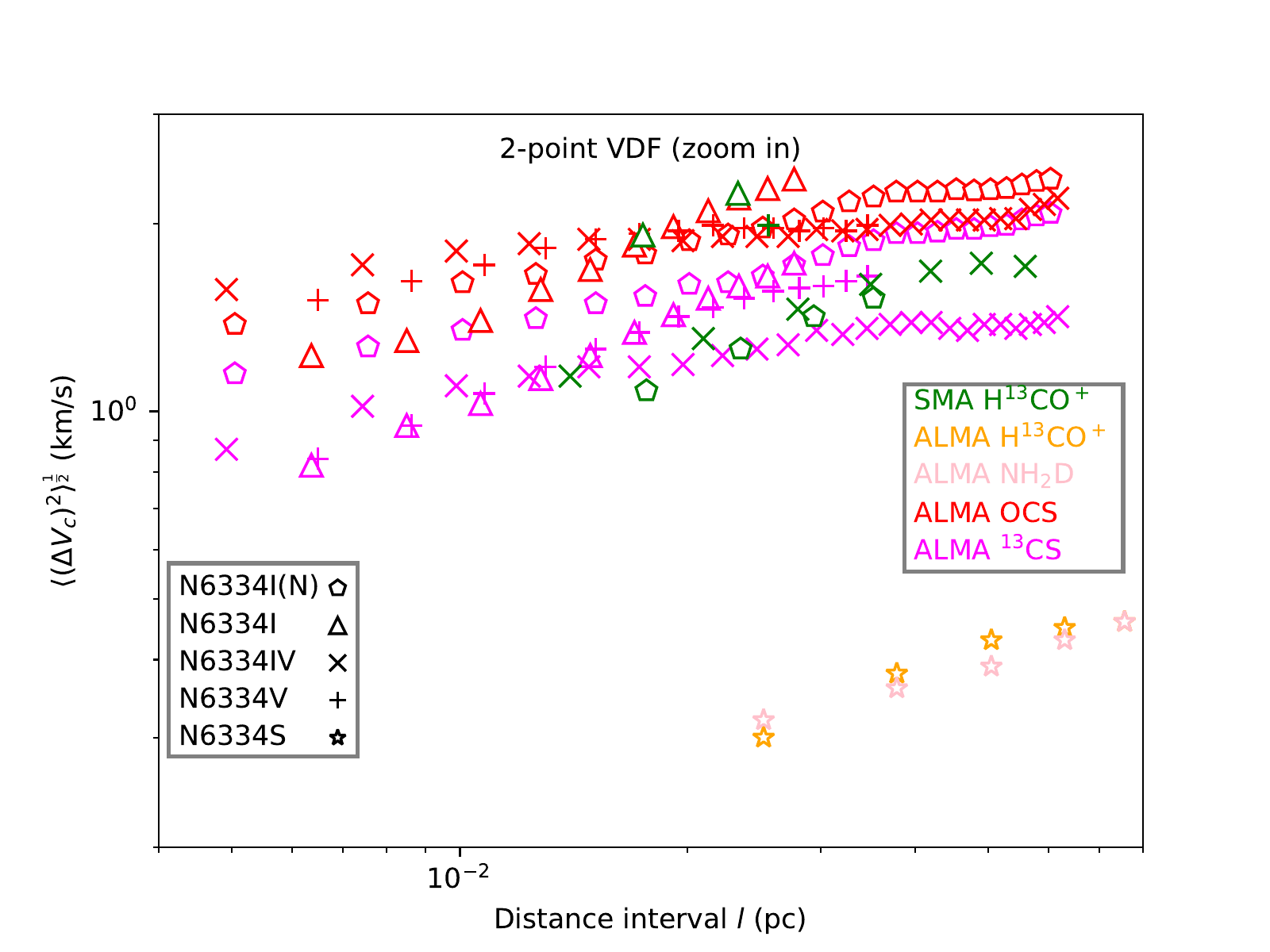}{0.48\textwidth}{}
 }
  \gridline{
 \fig{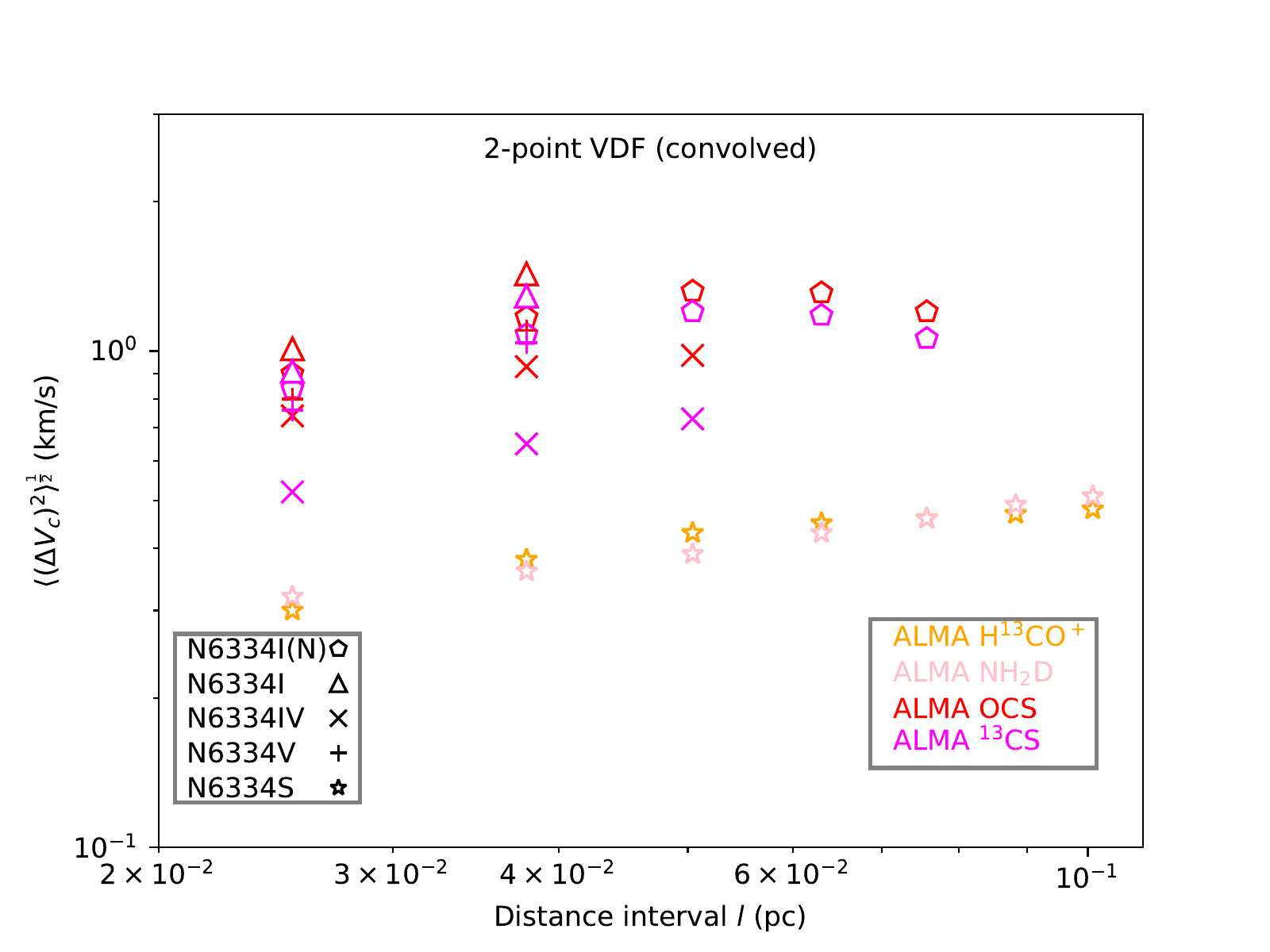}{0.48\textwidth}{}
 }
\caption{Upper left panel: 2-point dispersion functions of velocity centroids. The solid black line presents the result of a simple power-law fit for the JCMT VDF of N6334I(N) and the NANTEN2 VDF of the cloud. Upper right panel: a zoom-in of the upper left panel at small scales. Lower panel: convolved ALMA 2-point dispersion functions of velocity centroids. Different symbols represent different regions. Different colors indicate different instruments and lines. Only data points with $l$ greater than the beam resolution and smaller than the effective radius of the considered region are shown. \label{fig:N6334_vdf}}
\end{figure*}

We use the velocity dispersion function (VDF) in the form of structure functions of velocity centroids to characterize the multi-scale velocity fields in NGC 6334. We calculate the conventionally 2-point VDFs at different distance intervals ($l$) as 
\begin{equation}
\langle \Delta V_c^2(\boldsymbol{l}) \rangle^{\frac{1}{2}} = \langle( V_c(\boldsymbol{x}+\boldsymbol{l}) - V_c(\boldsymbol{x}))^2\rangle^{\frac{1}{2}}_x.
\end{equation}
The data are divided into separate distance bins with sizes corresponding to half of the beam size. For the NANTEN2 data, we only consider the area with NANTEN2 integrated $^{12}$CO (1-0) intensity greater than 25 K km s$^{-1}$ which approximately corresponds to the area of the NGC 6334 complex \citep{2018PASJ...70S..41F}. We consider the whole mapping area for the JCMT data. For the SMA and ALMA data, we only consider regions with S/N$>$3 for both the dust continuum and integrated line intensity (moment 0) maps (not shown in this paper). Because the small-scale statistics could be affected by the beam-smoothing effect \citep{1994ApJ...429..645M}, we only consider data points with $l \gtrsim l_\mathrm{beam}$. Because the large-scale statistics could be affected by the sparse sampling effect \citep{2022FrASS...9.3556L}, we only consider data points with distance intervals smaller than the effective radius $R_{eff}=\sqrt{A/\pi}$, where $A$ is the area of each considered region. Moreover, because of the large number of data points included in each distance bin, the statistical errors for the VDFs are negligible \citep{1994ApJ...429..645M}. Thus, we do not present error bars for the VDFs. 

The upper left panel of Figure \ref{fig:N6334_vdf} shows the VDFs for different lines, while the upper right panel presents a zoom-in of the VDFs at small scales. All the VDFs show positive slopes. Similar to the angular dispersion function (ADF) of polarization position angles \citep[see Figure 1 of][]{2009ApJ...696..567H}, the positive slopes of the VDFs are natural results of a combination of local turbulent and ordered velocity fields. The VDF of N6334I(N) at clump-to-core scales ($\sim$1-0.1 pc) revealed by JCMT $^{13}$CO (3-2) observations seems to roughly follow the same power-law as the VDF at cloud-to-clump scales ($\sim$10-1 pc) revealed by NANTEN2 $^{12}$CO (1-0) observations, which suggests that the non-thermal motions in N6334I(N) at clump-to-core scales mainly originate from the cascade of the cloud turbulence. With a simple least-square fit for the continuous power-law of the JCMT VDF of N6334I(N) and the NANTEN2 VDF of the cloud, we obtain $\langle(\Delta V_c)^2\rangle^{\frac{1}{2}} \sim l^{0.44}$ (see Figure \ref{fig:N6334_vdf}). The power-law index of 0.44 is consistent with previous VDF studies \citep[e.g., $\sim$0.43,][]{1994ApJ...429..645M}. On the other hand, the JCMT VDF of N6334I at clump-to-core scales is clearly higher (by a factor of $\sim$2) than the fitted power-law, which suggests that the clump-to-core scale non-thermal motions in N6334I have been increased by massive star formation activities. 

At sub-core ($\lesssim$0.1 pc) scales revealed by SMA and ALMA, the VDFs are comparable to or higher than the JCMT VDFs at clump-to-core scales and are much higher (up to a factor of $\sim$20) than the fitted power-law, suggesting that the non-thermal motions are significantly increased in small-scale high-density regions. For interferometric observations, the filtering of large-scale emissions could be a source of uncertainty for the VDF analysis. There is a lack of studies on how the filtering effect could affect the VDFs. If the non-thermal motions are universal at different densities and the VDFs behave similarly to the ADFs, the filtering effect should only affect data points at scales close to $l_{\mathrm{MRS}}$ \citep{2016ApJ...820...38H}. Thus, the filtering effect might not significantly affect our SMA and ALMA VDFs except for the largest several $l$ bins. This is because we only consider data points with $l$ smaller than the effective radius and we have $R_{eff}<l_{\mathrm{MRS}}$ for all the interferometer VDFs. However, due to the lack of theoretical studies, it is unclear whether the filtering effect should affect the VDFs and ADFs similarly. Also, it is expected that the non-thermal motions are more significant in higher-density regions at the same scale due to more active star formation activities. As interferometric observations filter out the diffuse emissions, the small-scale regions in diffuse gas materials with flat density profiles cannot be observed with interferometers. Thus, we cannot rule out the possibility that the velocity statistics in small-scale low-density regions still follow the large-scale turbulence cascade.


Other than the increase of VDFs in small-scale high-density regions, we also see that the VDFs from the same instrument and line but in different regions show different levels of amplitudes at the same scale. As discussed in Section \ref{sec:discussion} below, the increase of non-thermal motions in high-density regions could be related to massive star formation activities near young stellar objects (YSOs) such as infall, rotation, and/or HII regions. Small-scale outflow- or gravity-driven turbulence may also contribute to the increase of non-thermal motions. Thus, the varying evolutionary stages and star formation activities in different regions could naturally explain the different VDF amplitudes in different clumps. 

On the other hand, the VDFs from different lines but of the same clump show different amplitudes in N6334I(N), I, IV, and V. For the ALMA line observations toward these clumps, the OCS VDFs are always higher than the $^{13}$CS VDFs toward the same clump. This could be possibly explained by the higher $E_{\mathrm{u}}/k$ of the OCS (19-18) line. i.e., the OCS (19-18) line could trace hotter materials closer to the YSOs with more active star formation activities. The variations in the optical depth and chemical history of the two lines could also contribute to the difference in the VDFs. Similar reasons, adding the difference in critical densities, beam resolution, filtering scale, and area of the considered region, might explain the inconsistency between the SMA H$^{13}$CO$^+$ VDFs and the ALMA VDFs toward the same clumps. The situation is different in N6334S. Although the physical conditions traced by H$^{13}$CO$^+$ (1-0) and NH$_2$D (1$_{1,1}$-1$_{0,1}$) are different \citep{2020ApJ...896..110L}, the VDFs of the two lines are similar in shape and magnitude in N6334S. This may suggest that the non-thermal motions are coherent in different gas materials in N6334S due to the quiescent star formation activities. 

To compare the ALMA VDFs in different clumps and reduce the influence of the different beam resolutions, we convolve the ALMA line data toward N6334I(N), I, IV, and V to a similar resolution to N6334S. The VDFs for the convolved ALMA data are shown in the lower panel of Figure \ref{fig:N6334_vdf}. The ALMA H$^{13}$CO$^+$ and NH$_2$D VDFs toward the infrared dark clump N6334S are much lower than the convolved ALMA OCS and $^{13}$CS VDFs toward other more evolved clumps. Although this could be partially due to the difference in the filtering scale and line excitation condition, we suggest that the active star formation activity in more evolved clumps may be the main reason for the observed higher VDFs in N6334I(N), I, IV, and V. Future observational studies with the same line tracers and observational parameters toward regions at different evolutionary stages are necessary to further confirm this.


\subsubsection{Three-point velocity centroid dispersion functions}\label{sec:vdf3pt} 

\begin{figure*}[!htbp]
 \gridline{\fig{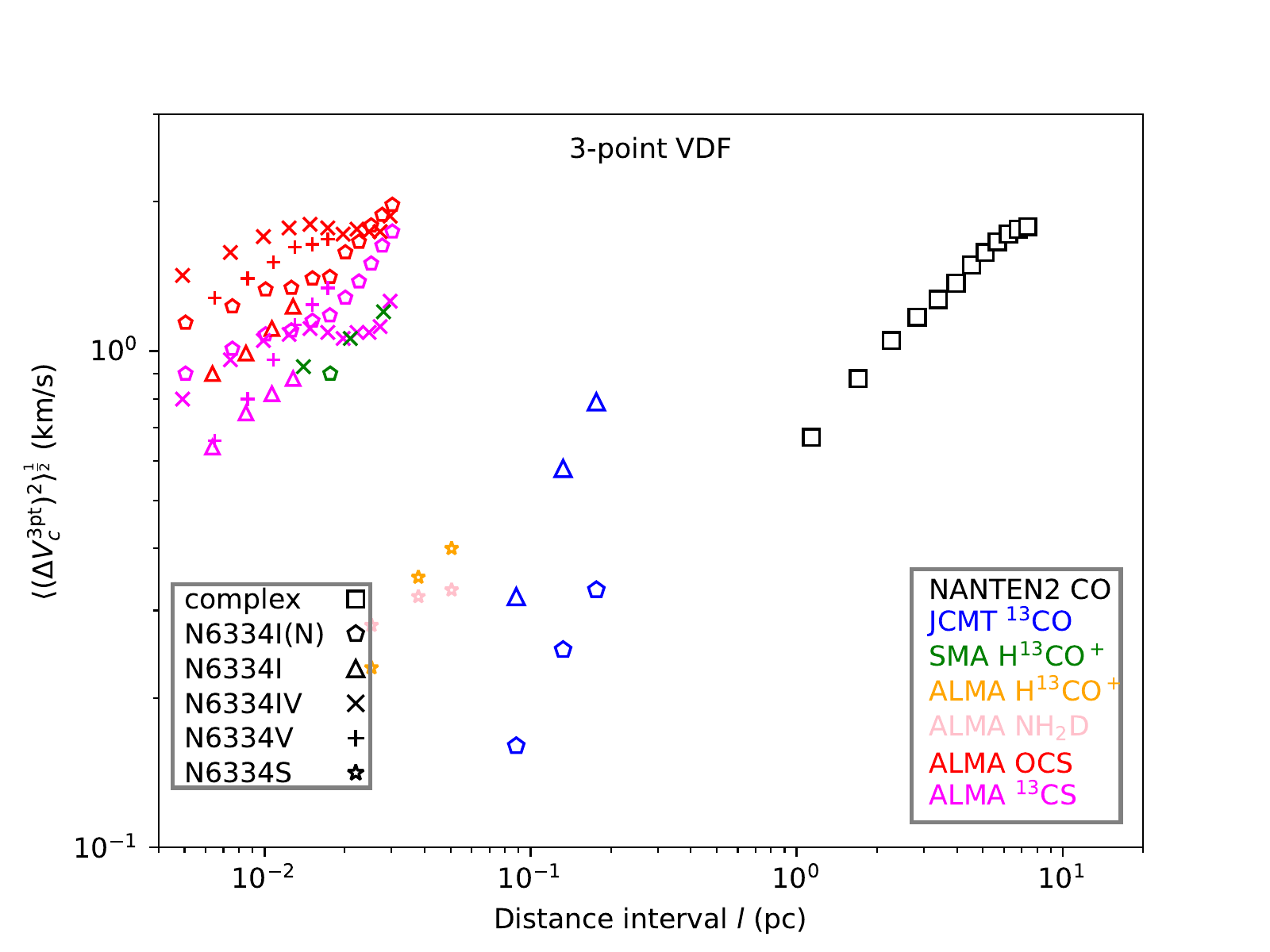}{0.48\textwidth}{}
 \fig{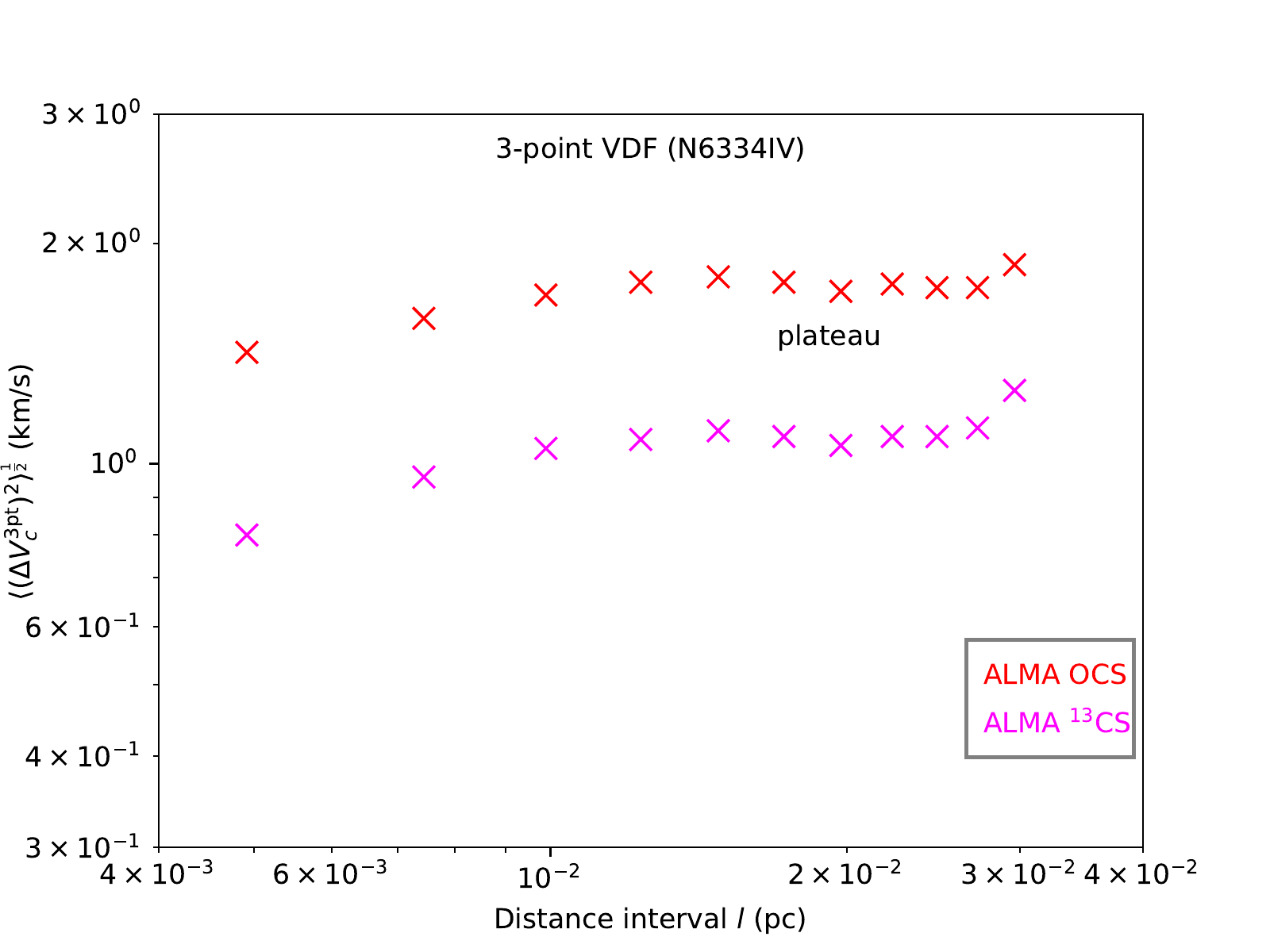}{0.48\textwidth}{}
 }
\caption{Left: 3-point dispersion functions of velocity centroids. Right: same as left but only for the ALMA VDF of N6334IV. Different symbols represent different regions. Different colors indicate different instruments and lines. Only data points with $l$ greater than the beam resolution and smaller than half of the effective radius of the considered region are shown. \label{fig:N6334_vdf3pt}}
\end{figure*}

The ordered velocity gradients at different scales in NGC 6334 (see Appendix \ref{sec:linem1}) make it difficult to derive pure turbulence statistics with the non-thermal VDFs. \citet{2022ApJ...935...77L} suggested that the ordered field contribution to the VDFs can be removed by calculating multi-point dispersion functions \citep{2019ApJ...874...75C} instead of the usually adopted 2-point dispersion functions. The simplest 3-point VDF is calculated as \citep{2019ApJ...874...75C}
\begin{equation}
\langle \Delta (V_c^{\mathrm{3pt}})^2(\boldsymbol{l}) \rangle^{\frac{1}{2}} = \langle( V_c(\boldsymbol{x}-\boldsymbol{l}) - 2V_c(\boldsymbol{x}) +V_c(\boldsymbol{x}+\boldsymbol{l}))^2\rangle^{\frac{1}{2}}_x/\sqrt{3}. 
\end{equation}
It is easily understood that the 3-point VDF can exactly remove a constant velocity gradient by averaging negative and positive gradients at distance $\boldsymbol{l}$. Similarly, an n-point VDF can exactly remove a large-scale $V_c(\boldsymbol{x})$ in the form of a polynomial of degree n-2 \citep{2019ApJ...874...75C}. Thus, multi-point dispersion functions can remove complex ordered fields by treating the ordered fields as high-order polynomials. Using simulated maps with small-scale fluctuations superimposed on large-scale variations, \citet{2019ApJ...874...75C} demonstrated that the multi-point dispersion function can effectively remove the large-scale variations and leave only the small-scale turbulent fluctuations. More-point dispersion functions can remove more complex ordered fields but will also increase the noise. If the ordered velocity field is successfully removed, the multi-point VDF should show a plateau corresponding to the level of the local turbulence \citep{2019ApJ...874...75C}. It should be noted that the multi-point dispersion function could remove the low spatial frequency turbulence as well as the ordered velocity fields \citep{2022FrASS...9.3556L}.

Following \citet{2019ApJ...874...75C}, we calculate the 3-point VDFs for the same molecular line data used for calculating the 2-point VDFs. The left panel of Figure \ref{fig:N6334_vdf3pt} shows the derived 3-point VDFs. The 3-point VDFs are lower and present larger scatters than the 2-point VDFs, but the relation between 3-point VDFs for different lines and regions is similar to those of the 2-point VDFs. On the other hand, except for the ALMA VDFs of N6334IV which show a clear plateau (right panel of Figure \ref{fig:N6334_vdf3pt}), the 3-point VDFs for most lines still show positive slopes without signs of plateau-like features. This is in agreement with the velocity centroid maps (Figure \ref{fig:N6334_alma_line_m1}) where N6334IV does not show clearly ordered velocity gradients. The positive slopes in other 3-point VDFs suggest that there are ordered velocity components remaining or that the upper $l$ limit of the VDF is smaller than the $l$ range of the plateau. Unfortunately, the size-to-resolution ratio of our data does not allow us to calculate more-point dispersion functions to further remove the contribution from ordered velocity fields. It is unclear how the filtering effect of interferometers could affect the multi-point dispersion function analysis, which needs to be addressed by future theoretical and numerical studies.

\subsection{Sonic Mach number} \label{sec:Ms}

\begin{figure*}[!htbp]
  \gridline{\fig{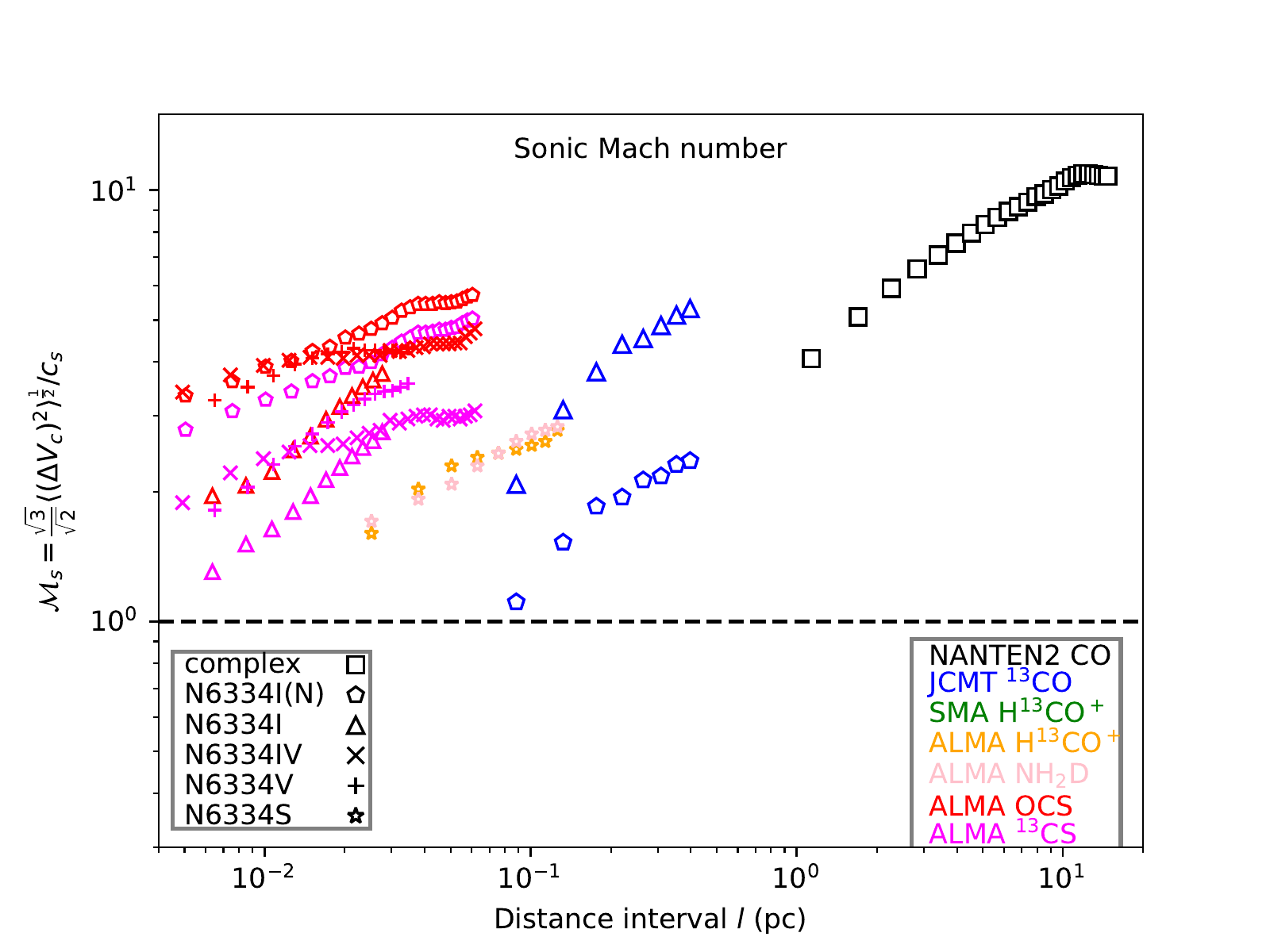}{0.48\textwidth}{}
 }
\caption{Sonic Mach number as a function of distance interval. Different symbols represent different regions. Different colors indicate different instruments and lines. Only data points with $l$ greater than the beam resolution and smaller than the effective radius of the considered region are shown. \label{fig:N6334_Ms}}
\end{figure*}

We estimate the sonic Mach number for the non-thermal motions with the 2-point VDFs. Similar to the relation between the ADF and the angular dispersion \citep{2009ApJ...696..567H}, the VDF and the velocity centroid dispersion ($\delta V_c$) have the relation $\langle(\Delta V_c)^2\rangle^{\frac{1}{2}} = \sqrt{2(\langle V_c^2\rangle - \langle V_c\rangle^2)} = \sqrt{2}\delta V_c$. Assuming that $\delta V_c$ is $1/\sqrt{3}$ times of the 3D non-thermal velocity dispersion, we estimate the 3D sonic Mach number\footnote{Some previous studies report the 1D sonic Mach number, which is the ratio between the 1D non-thermal velocity dispersion and sound speed. There is a factor of $\sqrt{3}$ difference between the 1D and 3D sonic Mach number.} with $\mathcal{M}_s = \sqrt{3} \langle(\Delta V_c)^2\rangle^{\frac{1}{2}}/(\sqrt{2}c_s)$. The 1D sound speed for isothermal gas is estimated as $c_s = \sqrt{k_\mathrm{B}T/(\mu_\mathrm{p}m_\mathrm{H})}$, where $k_\mathrm{B}$ is the Boltzmann constant, $T$ is the gas temperature, $\mu_\mathrm{p}=2.37$ \citep{2008A&A...487..993K} is the mean molecular weight per free particle, and $m_\mathrm{H}$ is the atomic mass of hydrogen. We adopt $T=20$ K for NANTEN2 and JCMT observations because the temperature variation in NGC 6334 is small at these scales \citep{2021A&A...647A..78A}. We adopt $T=15$ K for ALMA observations toward N6334S \citep{2020ApJ...896..110L}. We do not calculate $\mathcal{M}_s$ for the SMA observations due to the lack of temperature estimation. The average temperature of our considered regions in N6334I(N), I, IV, and V are $\sim$61, 167, 67, and 83 K, respectively, for the ALMA observations \citep{2023ApJ...945..160L}. Figure \ref{fig:N6334_Ms} shows the calculated $\mathcal{M}_s$ at different scales. The derived $\mathcal{M}_s$ ranges from $\sim$11 to $\sim$1.1 at $\sim$15-0.005 pc scales, which stays supersonic\footnote{Some previous studies adopt $\mathcal{M}_s>1$ as supersonic \citep[e.g., ][]{2021NatAs...5..365F}, while some other studies adopt $\mathcal{M}_s>2$ as supersonic \citep[e.g.,][]{2020ApJ...896..110L}. We adopt the former definition in this paper.} (i.e., $\mathcal{M}_s>1$) across 3-4 orders of magnitude. We also estimated the $\mathcal{M}_s$ for the small-scale turbulent motions in N6334IV with the 3-point ALMA VDFs, which stay supersonic as well. 

\section{Discussion} \label{sec:discussion}

\subsection{Non-thermal motions in massive star formation region}\label{sec:nt}

Non-thermal motions can be divided into non-turbulent (ordered) and turbulent motions. If both the ordered and turbulent motions originate from the same turbulence cascade, the small-scale ordered motion is actually part of the turbulent motion at larger scales \citep{1994ApJ...429..645M}, and the velocity statistics for non-thermal motions should show a continuous power-law relation across different scales. However, except for the JCMT VDF of N6334I(N) and the NANTEN2 VDF of the cloud, the VDFs from different datasets at different scales are not continuous in general (see Figure \ref{fig:N6334_vdf}). On the other hand, the VDFs from the same line toward different clumps and the VDFs from different lines toward the same clump show distinct magnitudes at the same scale, so the VDFs are not universal either. Thus, it is unlikely that the non-thermal motions in star formation regions in NGC 6334 are solely due to the cascade of global supersonic turbulence in the interstellar medium (ISM). It should be noted that the multi-scale non-thermal motions in low-density regions without star formation activities may still follow a continuous turbulence cascade. But such low-density regions are not the focus of this star formation study. 

Alternatively, it is expected that the massive star formation activity near YSOs can increase the non-thermal motions in small-scale and high-density regions as the region evolves. Such effects have been previously observed in massive star formation regions \citep[e.g.,][]{1998ApJ...505L.151Z}. The various star formation activities in different clumps can naturally explain the different magnitudes of VDFs at the same scale. Especially, the ALMA VDFs in the infrared dark N6334S are much lower than the ALMA VDFs in more evolved clumps at the same scale, which confirms that the non-thermal motion is increased as the clump/core evolves. This trend is in agreement with previous linewidth observations in the infrared dark cloud G28.34 \citep{2008ApJ...672L..33W}, in which they found that the large-scale linewidth is broader than the small-scale linewidth in an infrared dark clump (P1 or MM4) at an earlier evolution stage but is narrower than the small-scale linewidth in an infrared bright clump (P2 or MM1) that is more evolved. On the other hand, the ALMA VDFs for N6334S are comparable to or higher than the JCMT VDF toward N6334I(N) at a larger scale, which suggests the star formation activities have already increased the non-thermal motions in the early stages of massive star formation in massive dense cores in infrared dark regions. 

An interesting topic is the scale below which the non-thermal motion deviates from a continuous power-law. As revealed in Section \ref{sec:vdf2pt}, the JCMT VDF of N6334I(N) follows the same power-law relation as the NANTEN2 VDF of the cloud (see Figure \ref{fig:N6334_vdf}), while the JCMT VDF of N6334I is higher than this power-law. This suggests that the non-thermal motions in some star formation regions have already been increased at clump-to-core scales. Below $\sim$0.1 pc, all the VDFs are much higher than the fitted power-law, suggesting a significant increase of non-thermal motions at sub-core scales. On the other hand, we cannot rule out the possibility that the velocity statistics in low-density regions without star formation activities could still follow the large-scale power-law relation even at sub-core scales. Thus, we conservatively conclude that with the given data we cannot identify a specific scale below which the velocity statistics deviate from a power-law relation.

\subsection{Turbulent motions in massive star formation region}\label{sec:turb}

Non-driven supersonic turbulence decays quickly, thus inputs of energy must exist to maintain the turbulence within star-forming molecular clouds \citep{2004RvMP...76..125M}. It has been long debated whether the turbulence is driven at large-scale \citep[e.g., $\sim$100 pc by supernovae, ][]{2004ApJ...617..339B} and cascades to small-scales, or whether the turbulence is driven by small-scale mechanisms such as outflows \citep{2006ApJ...640L.187L, 2007ApJ...662..395N} or gravitational accretion \citep{2010A&A...520A..17K}. Observational studies in low-mass star formation regions usually found a power-law relation for the velocity statistics at different scales \citep[e.g., Taurus,][]{2022arXiv220413760Y}, which provide evidence for a continuous turbulent cascade. However, there is a lack of observational studies of the turbulence cascade in massive star formation regions where the gravitational effect and protostellar feedback are more significant. 

In sites of massive star formation, there could be significant amounts of non-turbulent non-thermal motions (e.g., infall or rotation) in small-scale and high-density regions \citep{2021ApJ...919...79L}, which makes it difficult to derive pure turbulence statistics that are required for some statistical analyses \citep[e.g., the DCF analysis;][]{2022ApJ...925...30L}. We make the attempt to remove the ordered velocity field contribution by calculating the 3-point VDFs, but only the ALMA 3-point VDF for N6334IV shows a plateau that is expected for pure turbulence. The plateau of the 3-point ALMA VDFs for N6334IV is comparable to or higher than the NANTEN2 and JCMT VDFs at larger scales, which may provide observational evidence for turbulence driven by small-scale mechanisms in massive star formation regions. Such small-scale-driven turbulence could potentially reduce the star formation rate, affect the small-scale fragmentation, and shift the initial mass function to lower masses \citep{2021MNRAS.507.2448M, 2022MNRAS.513.2100H}. 

\section{Summary} \label{sec:summary}
We implement 2-point and 3-point velocity centroid structure functions on the NANTEN2 $^{12}$CO (1-0), JCMT $^{13}$CO (3-2), SMA H$^{13}$CO$^+$ (4-3), and ALMA H$^{13}$CO$^+$ (1-0), NH$_2$D (1$_{1,1}$-1$_{0,1}$), OCS (19-18), and $^{13}$CS (5-4) line data to study the multi-scale statistics of the non-thermal motions in the massive star formation region NGC 6334. The findings are:
\begin{enumerate}
    \item The combination of small-scale turbulent motions and ordered velocity fields induced by massive star formation activities increases the non-thermal motion in small-scale and high-density star formation regions, which makes the multi-scale non-thermal motions in active star formation regions deviate from a continuous turbulence cascade.
    \item The non-thermal motion revealed by the same line varies in different clumps at the same scale, which is due to the different star formation activities in these clumps. The non-thermal motions revealed by different lines show different levels in the same clump at the same scale, which is due to the different gas materials traced by these line data. These differences show that the non-thermal motions of gas materials at different densities or in different star-forming subregions do not follow an universal turbulence cascade when star formation activities are strong within the molecular cloud. 
    \item The level of non-thermal motion in the infrared dark clump N6334S is lower than that of other more evolved clumps, confirming that the non-thermal motion is increased as clump/core evolves. 
    \item The 3-point ALMA VDFs in N6334IV show signs of pure turbulence statistics, which may provide observational evidences for small-scale-driven supersonic turbulence. 
    \item The non-thermal motion in NGC 6334 stays supersonic over several orders of magnitudes. 
\end{enumerate}


\begin{acknowledgments}
We thank the anonymous referee for the detailed comments. We thank Dr.Doris Arzoumanian for sharing the JCMT dust continuum maps and the Herschel temperature maps. We thank Prof. Yasuo Fukui and Dr.Mikito Kohno for sharing the NANTEN2 data. 

J.L. acknowledges the support from the EAO Fellowship Program under the umbrella of the East Asia Core Observatories Association. K.Q. is supported by National Key R\&D Program of China grant No. 2022YFA1603100. K.Q. acknowledges the support from National Natural Science Foundation of China (NSFC) through grant Nos. U1731237, 11590781, and 11629302. H.B.L. is supported by the Ministry of Science and Technology (MoST) of Taiwan (Grant Nos. 108-2112-M-001-002-MY3, 108-2923-M-001-006-MY3, 111-2112-M-001-089-MY3). ZYL is supported in part by NSF AST-1815784 and NASA 20NSSC18K1095. This work was also partially supported by the program Unidad de Excelencia María de Maeztu CEX2020-001058-M. JMG also acknowledges support by the grant PID2020-117710GB-I00 (MCI-AEI-FEDER, UE).

This paper makes use of the following ALMA data: ADS/JAO.ALMA\#2017.1.00793.S and ADS/JAO.ALMA\#2016.1.00951.S. ALMA is a partnership of the ESO (representing its member states), NSF (USA) and NINS (Japan), together with NRC (Canada), MOST and ASIAA (Taiwan), and KASI (Republic of Korea), in cooperation with the Republic of Chile. The Joint ALMA Observatory is operated by ESO, AUI/NRAO, and NAOJ. The National Radio Astronomy Observatory is a facility of the National Science Foundation operated under cooperative agreement by Associated Universities, Inc.
The JCMT is operated by the EAO on behalf of NAOJ; ASIAA; KASI; CAMS as well as the National Key R\&D Program of China (No. 2017YFA0402700). Additional funding support is provided by the STFC and participating universities in the UK and Canada. Additional funds for the construction of SCUBA-2 were provided by the Canada Foundation for Innovation.
This work is based on observations obtained with Planck (http://www.esa.int/Planck), an ESA science mission with instruments and contributions directly funded by ESA Member States, NASA, and Canada. 
The present study has also made use of NANTEN2 data. NANTEN2 is an international collaboration of ten universities: Nagoya University, Osaka Prefecture University, University of Cologne, University of Bonn, Seoul National University, University of Chile, University of New South Wales, Macquarie University, University of Sydney, and Zurich Technical University. 
\end{acknowledgments}

%

\vspace{5mm}
\facilities{NANTEN2, JCMT(HARP), SMA, ALMA}


\software{Astropy \citep{2013A&A...558A..33A,2018AJ....156..123A},  
Matplotlib \citep{2007CSE.....9...90H}.
          }

\appendix
\section{Velocity centroid maps}\label{sec:linem1}

\begin{figure*}[!htbp]
 \gridline{
 \fig{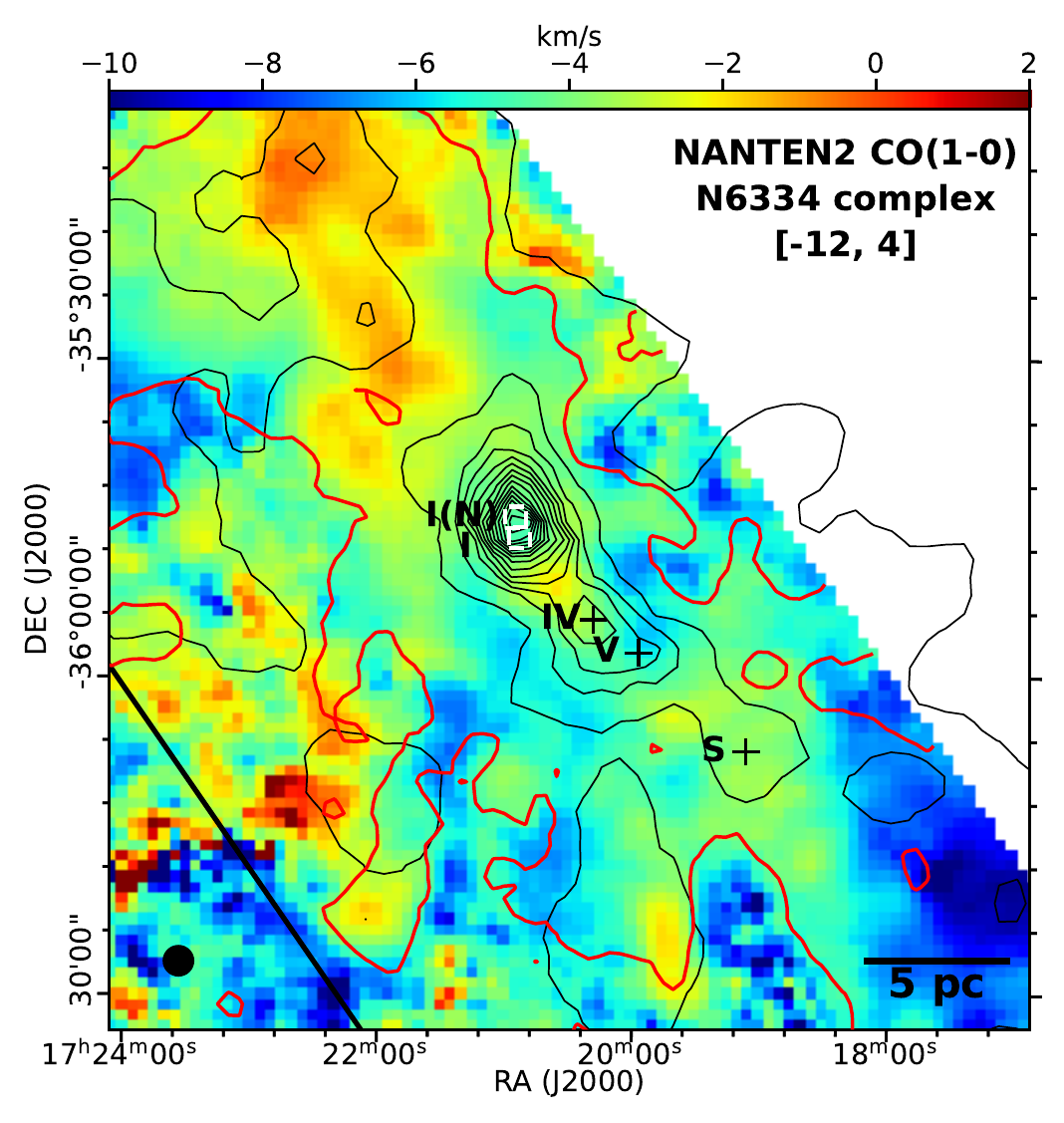}{0.33\textwidth}{(a)}
 \fig{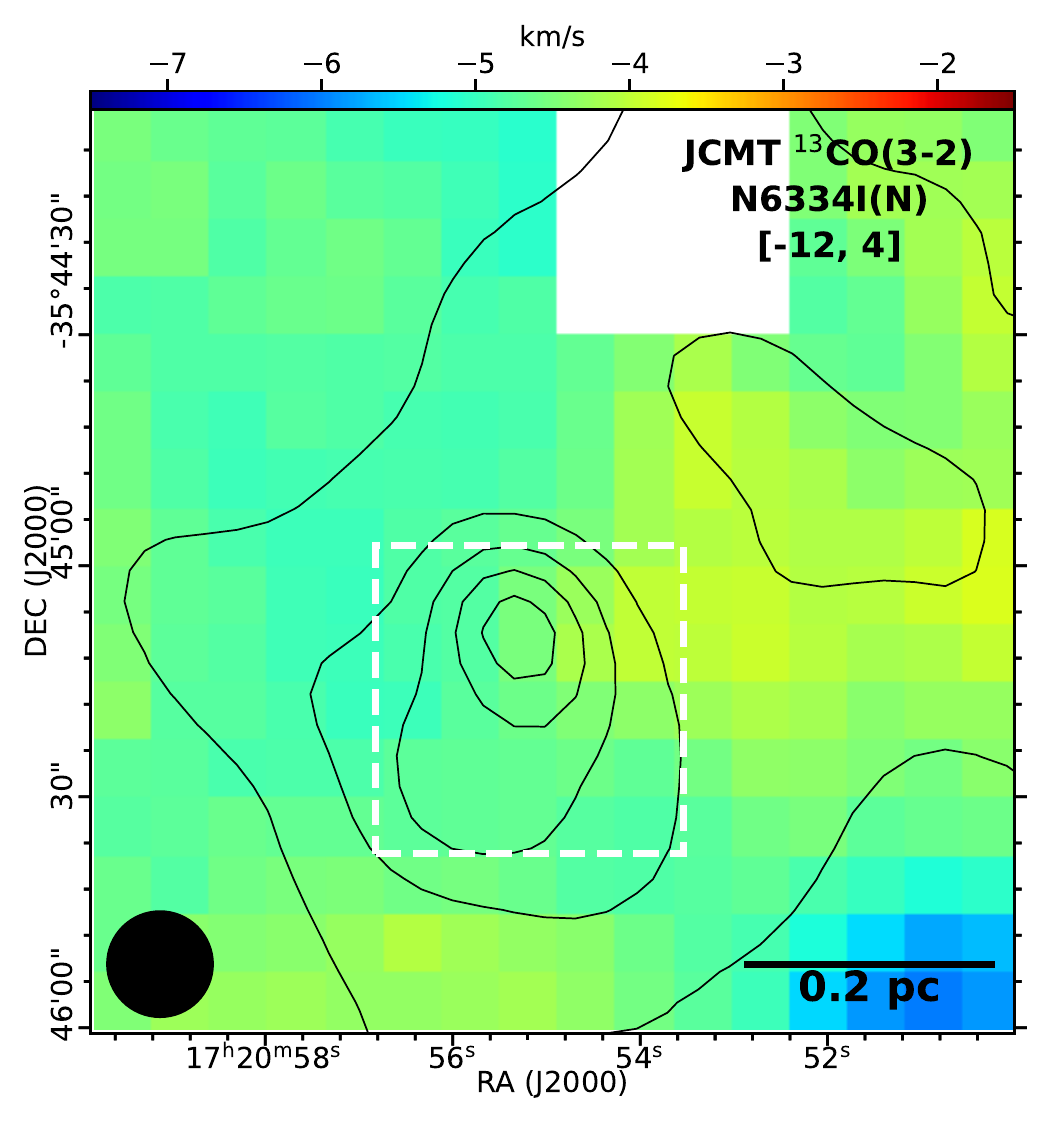}{0.33\textwidth}{(b)}
 \fig{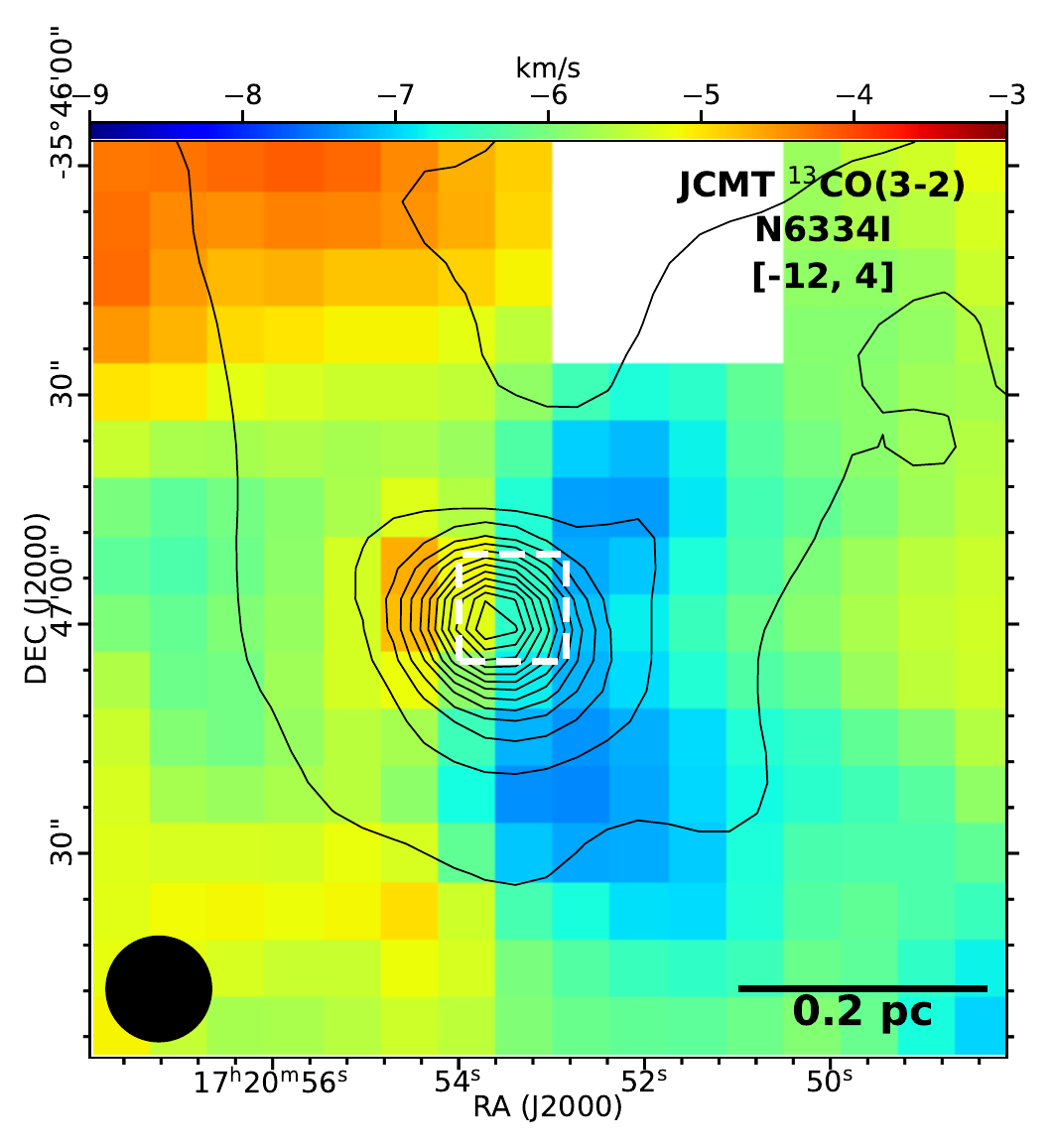}{0.33\textwidth}{(c)}
 }
\caption{(a) Velocity centroid map of NANTEN2 $^{12}$CO (1-0) line emission toward NGC 6334 complex \citep{2018PASJ...70S..41F}. The black contour levels correspond to the Planck 353 GHz (0.85 mm) optical depth ($\tau_{353}$) map \citep{2014A&A...571A..11P}. The contour starts at 0.0004 and continues with an interval of 0.0004. White rectangles indicate the map area of the JCMT fields in (b) and (c). Black crosses indicate the positions of N6334IV, V, and S. The red contour indicates the region within which we perform the VDF analysis. The galactic plane (black line) is indicated in the lower left corner. (b)-(c) Velocity centroid maps of JCMT $^{13}$CO (3-2) line emission toward N6334I(N) and N6334I. The black contour levels correspond to the JCMT 0.85 mm dust continuum map \citep{2021A&A...647A..78A}. Contour starts at 2 Jy beam$^{-1}$ and continues with an interval of 4 Jy beam$^{-1}$. White rectangles indicate the SMA and ALMA map area toward N6334I(N) and I in Figures \ref{fig:N6334_sma_line_m1} and \ref{fig:N6334_alma_line_m1}. The beam and a scale bar are indicated in the lower left and lower right corners of each panel, respectively. Panels (b) and (c) were previously shown in \citet{2023ApJ...945..160L} and are reproduced with permission.  \label{fig:N6334_large_line_m1}}
\end{figure*}

\begin{figure*}[!htbp]
 \gridline{
 \fig{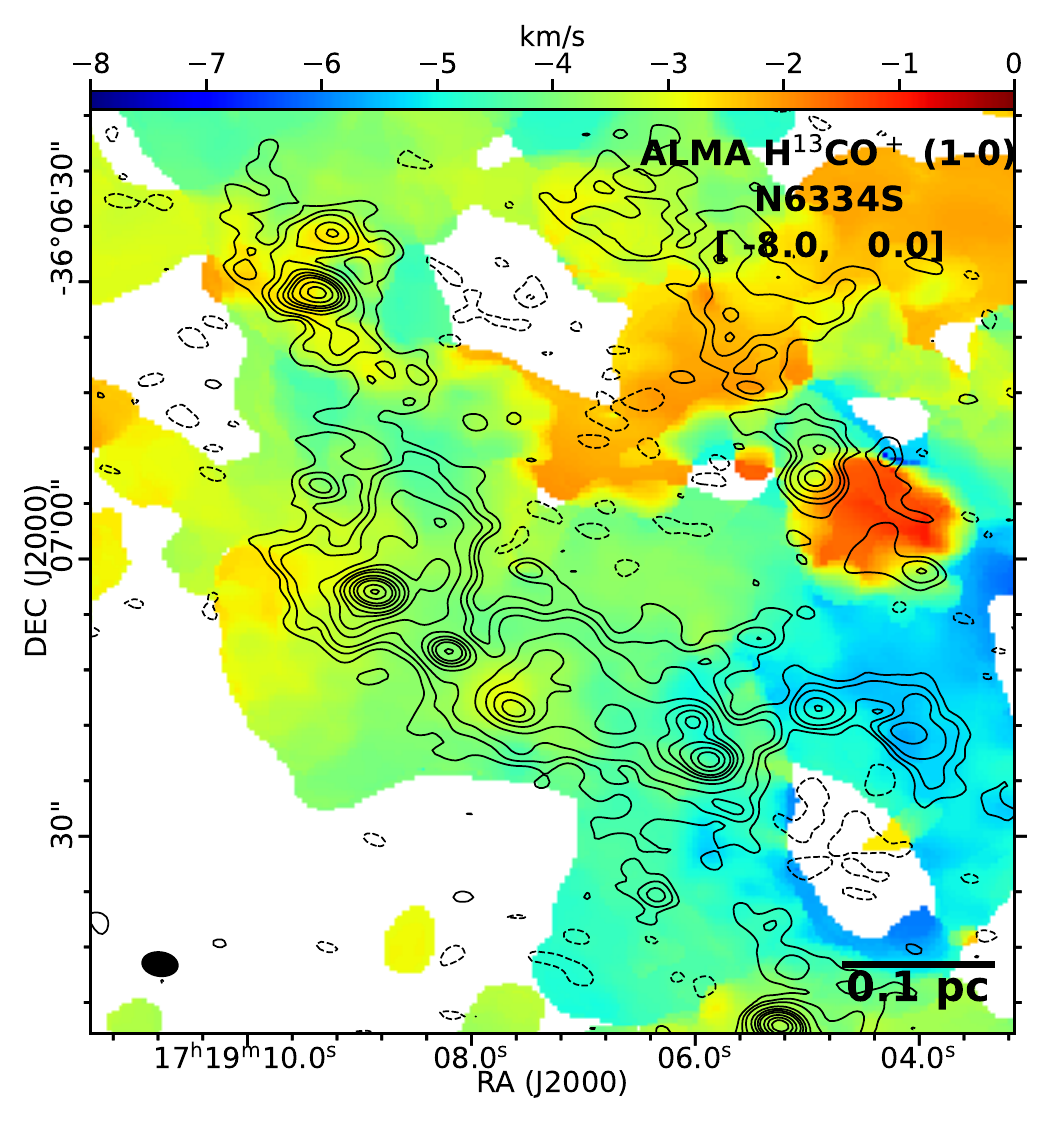}{0.33\textwidth}{(a)}
 \fig{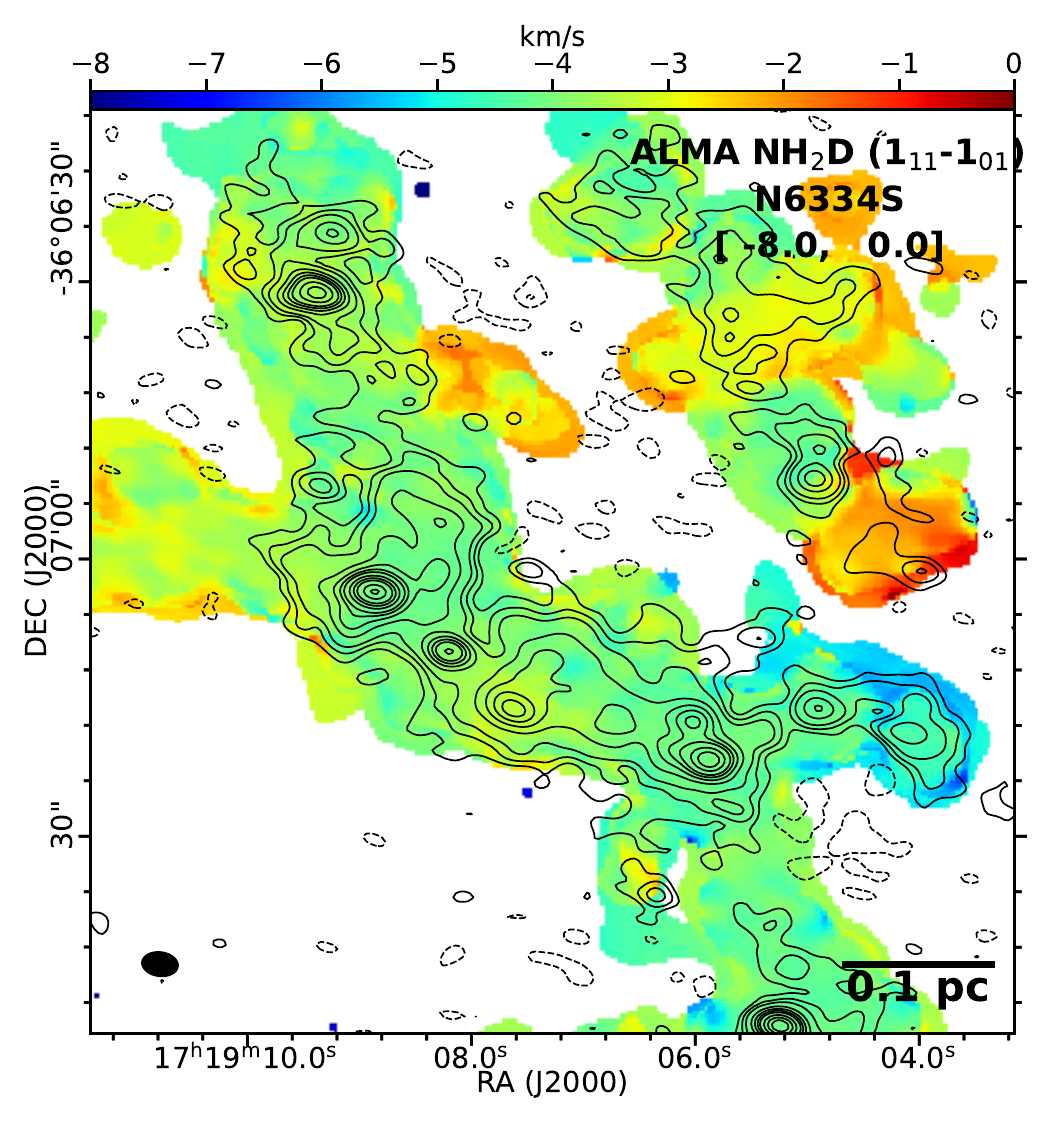}{0.33\textwidth}{(b)}
 }

\caption{Velocity centroid map of ALMA (a) H$^{13}$CO$^+$ (1-0) and (b) NH$_2$D (1$_{1,1}$-1$_{0,1}$) line emission toward N6334S \citep{2020ApJ...896..110L, 2022ApJ...926..165L}. The contour levels correspond to the ALMA 3 mm dust continuum map. Contour levels are ($\pm$3, 6, 10, 20, 30, 40, 50, 70, 90, 110, 150, 180, 210, 250, 290, 340, 390, 450) $\times \sigma_{I}$, where $\sigma_{I}=0.03$ mJy beam$^{-1}$ \citep{2020ApJ...896..110L} is the RMS noise of the Stokes $I$ map. The synthesized beam and a scale bar are indicated in the lower left and lower right corners of each panel, respectively. \label{fig:N6334_alma_S_line_m1}}
\end{figure*}

\begin{figure*}[!htbp]
 \gridline{
 \fig{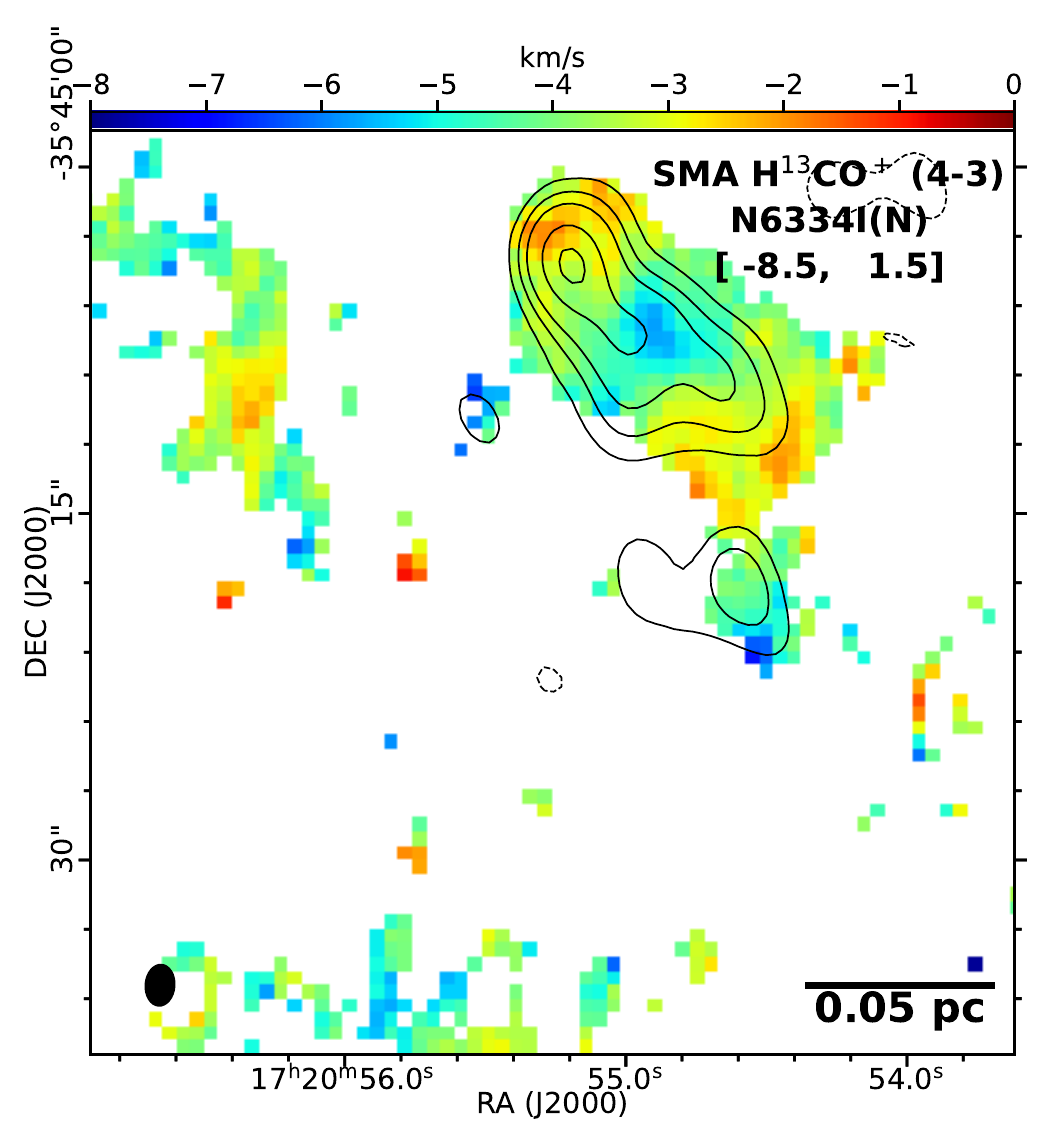}{0.33\textwidth}{(a)}
 \fig{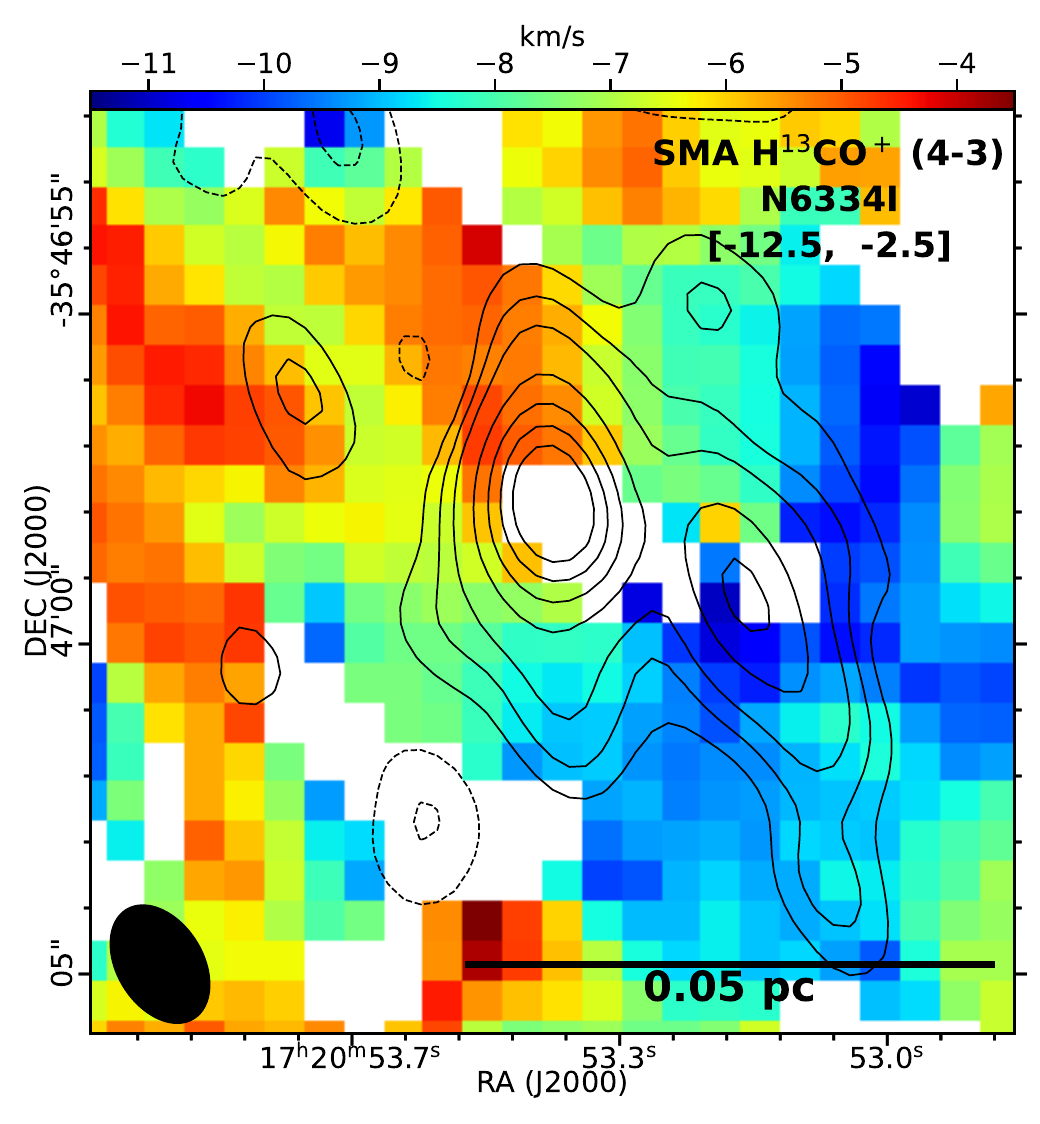}{0.33\textwidth}{(b)}
 }
 \gridline{
 \fig{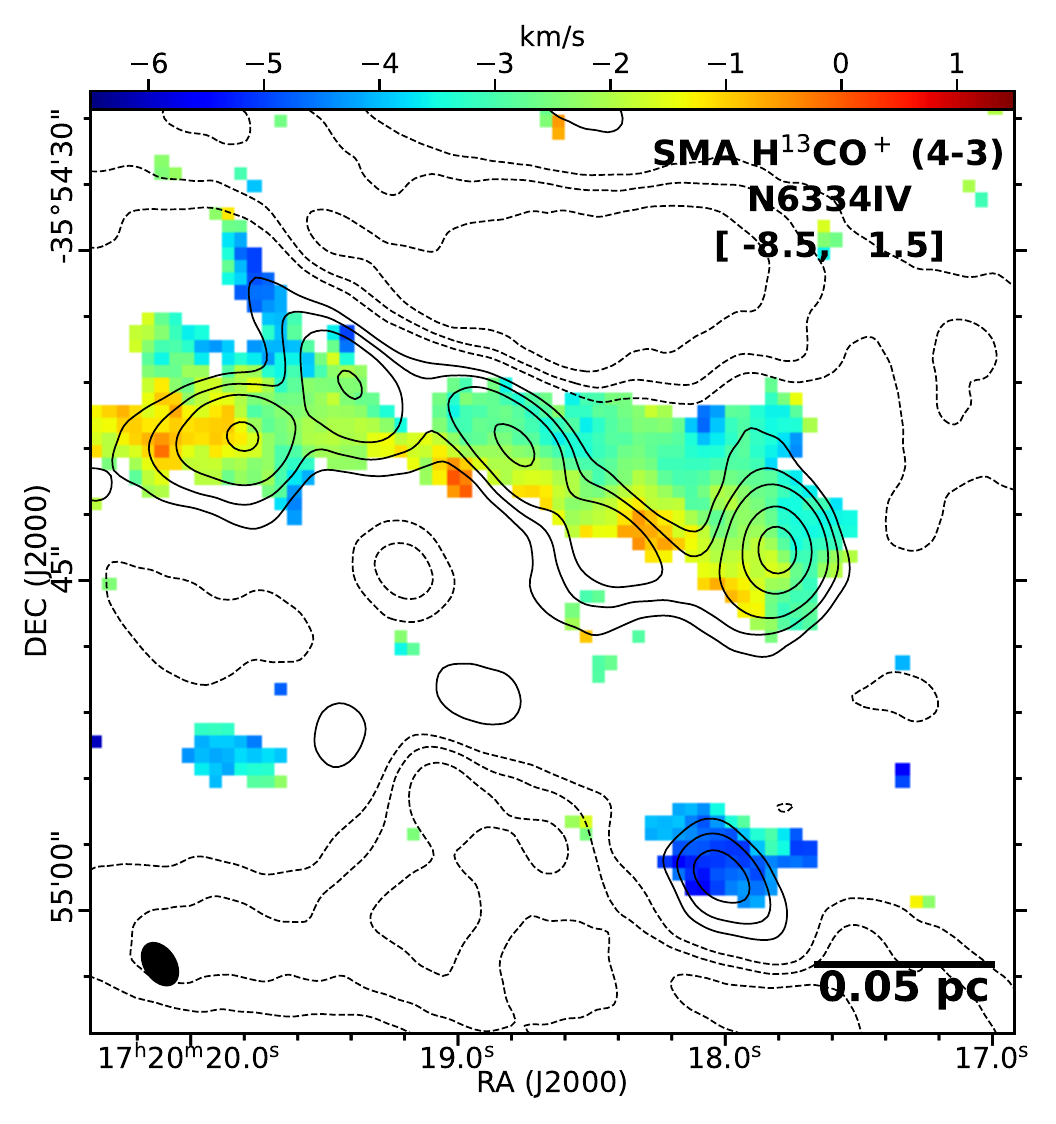}{0.33\textwidth}{(c)}
 \fig{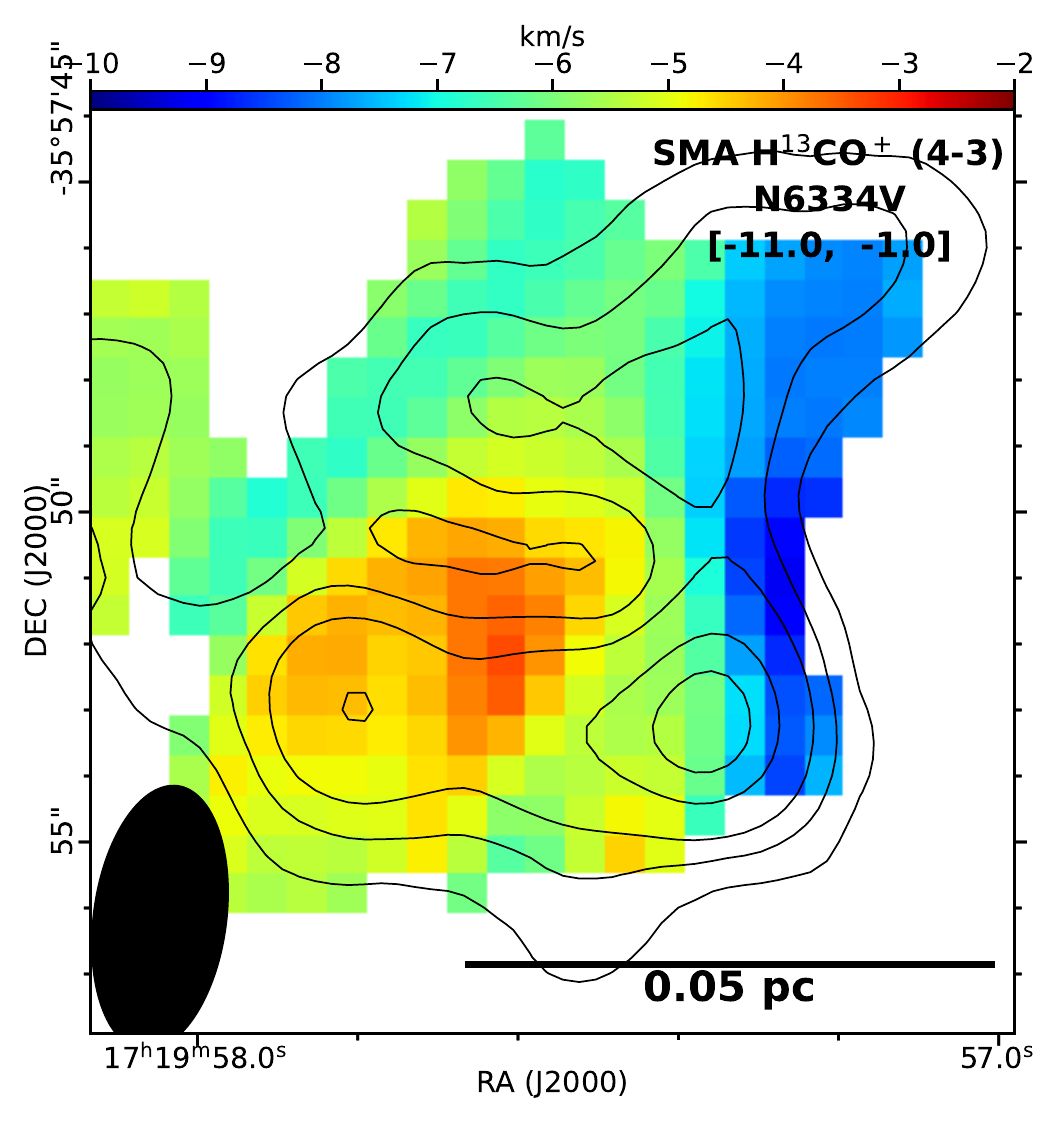}{0.33\textwidth}{(d)}
 }
\caption{Velocity centroid maps of SMA H$^{13}$CO$^+$ (4-3) observations. The contour levels correspond to the SMA 0.87 mm dust continuum map \citep{2014ApJ...792..116Z, 2021ApJ...912..159P}. Contour levels are ($\pm$3, 6, 10, 20, 30, 40, 50, 70, 90, 110, 150, 180, 210, 250, 290, 340, 390, 450) $\times \sigma_{I}$, where $\sigma_{I}=$100, 100, 30, and 30 mJy beam$^{-1}$ for N6334I(N), I, IV, and V, respectively. The synthesized beam and a scale bar are indicated in the lower left and lower right corners of each panel, respectively. \label{fig:N6334_sma_line_m1}}
\end{figure*}

\begin{figure*}[!htbp]
  \gridline{\fig{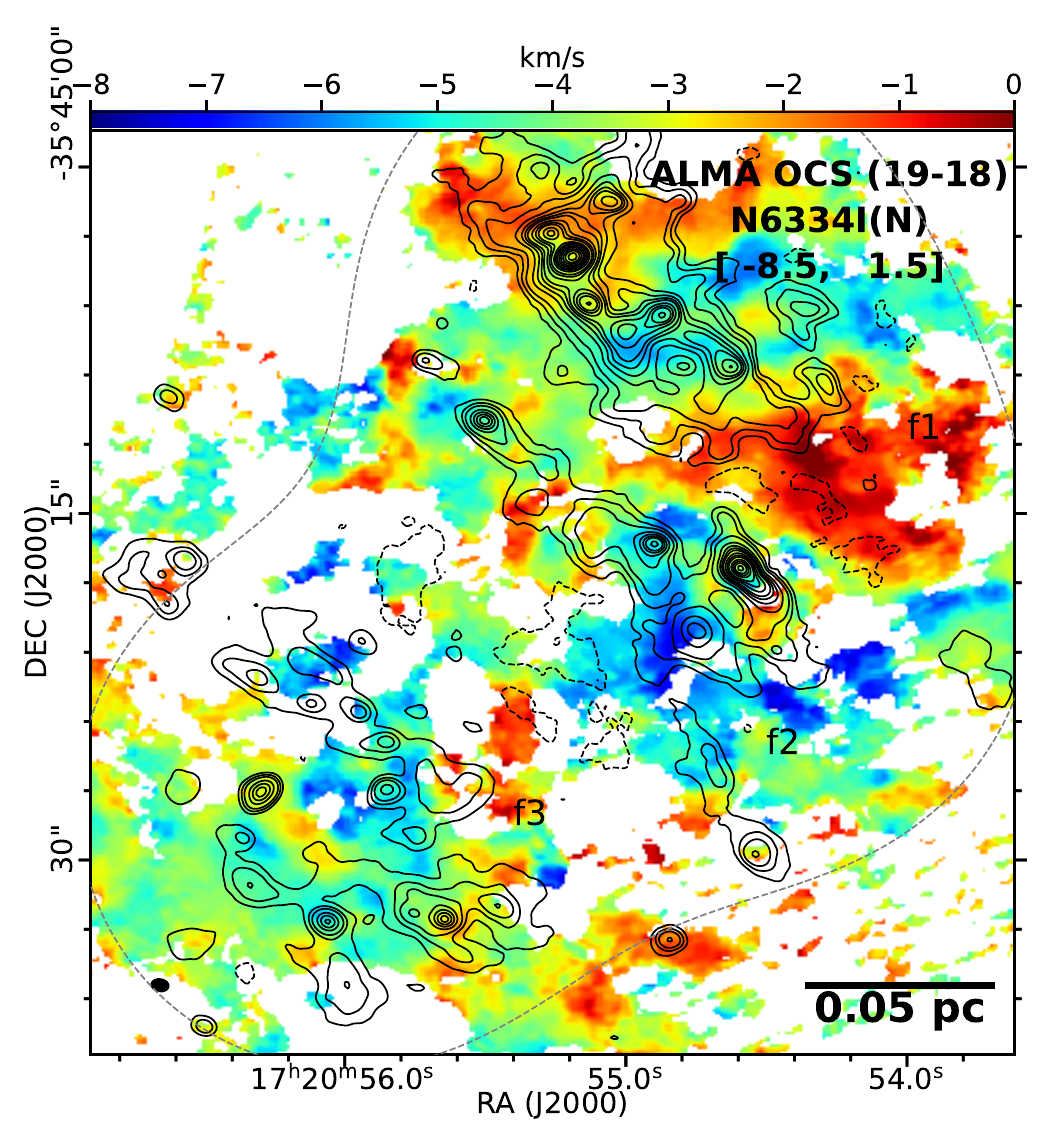}{0.33\textwidth}{(a)}
 \fig{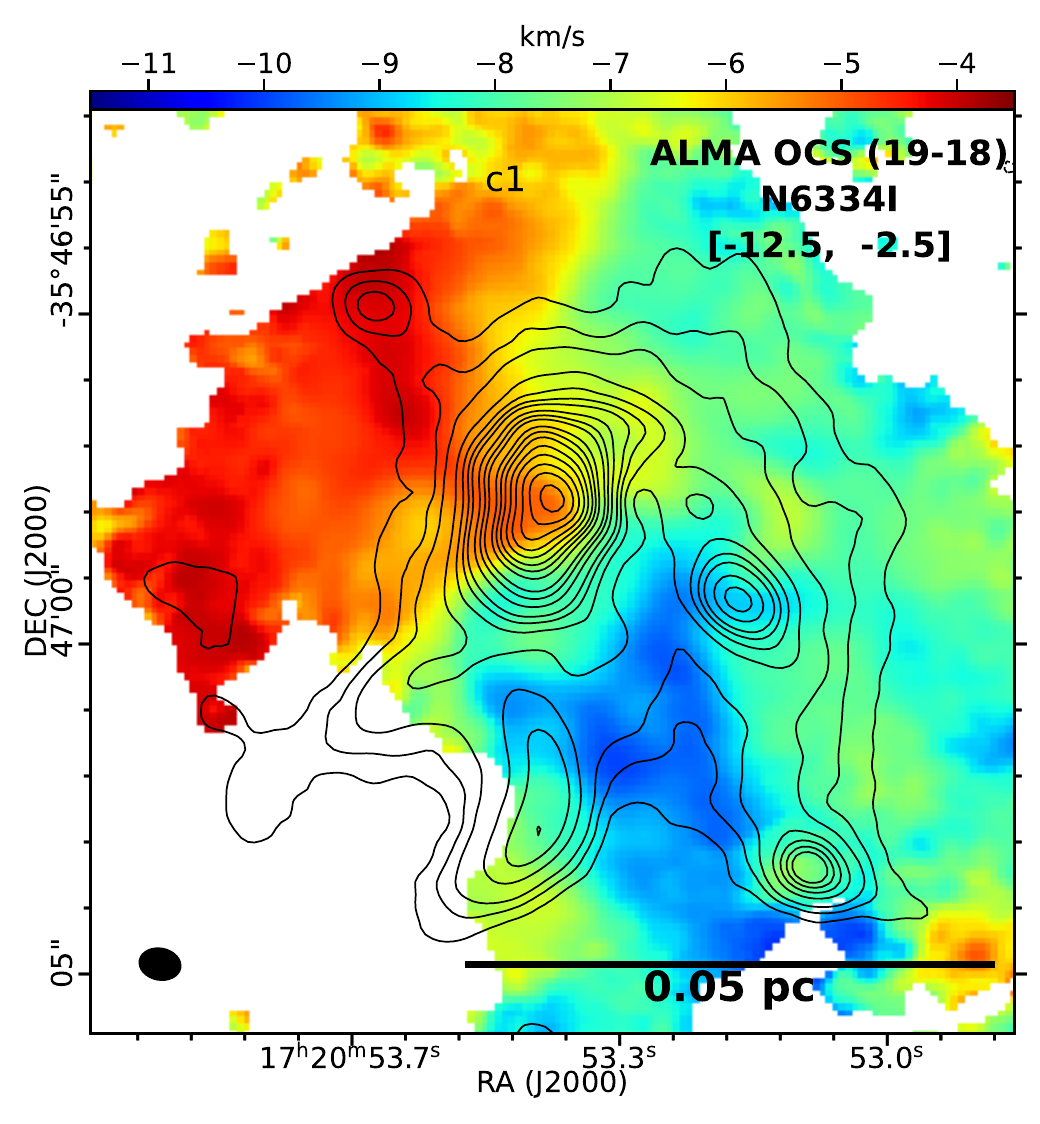}{0.33\textwidth}{(b)}
\fig{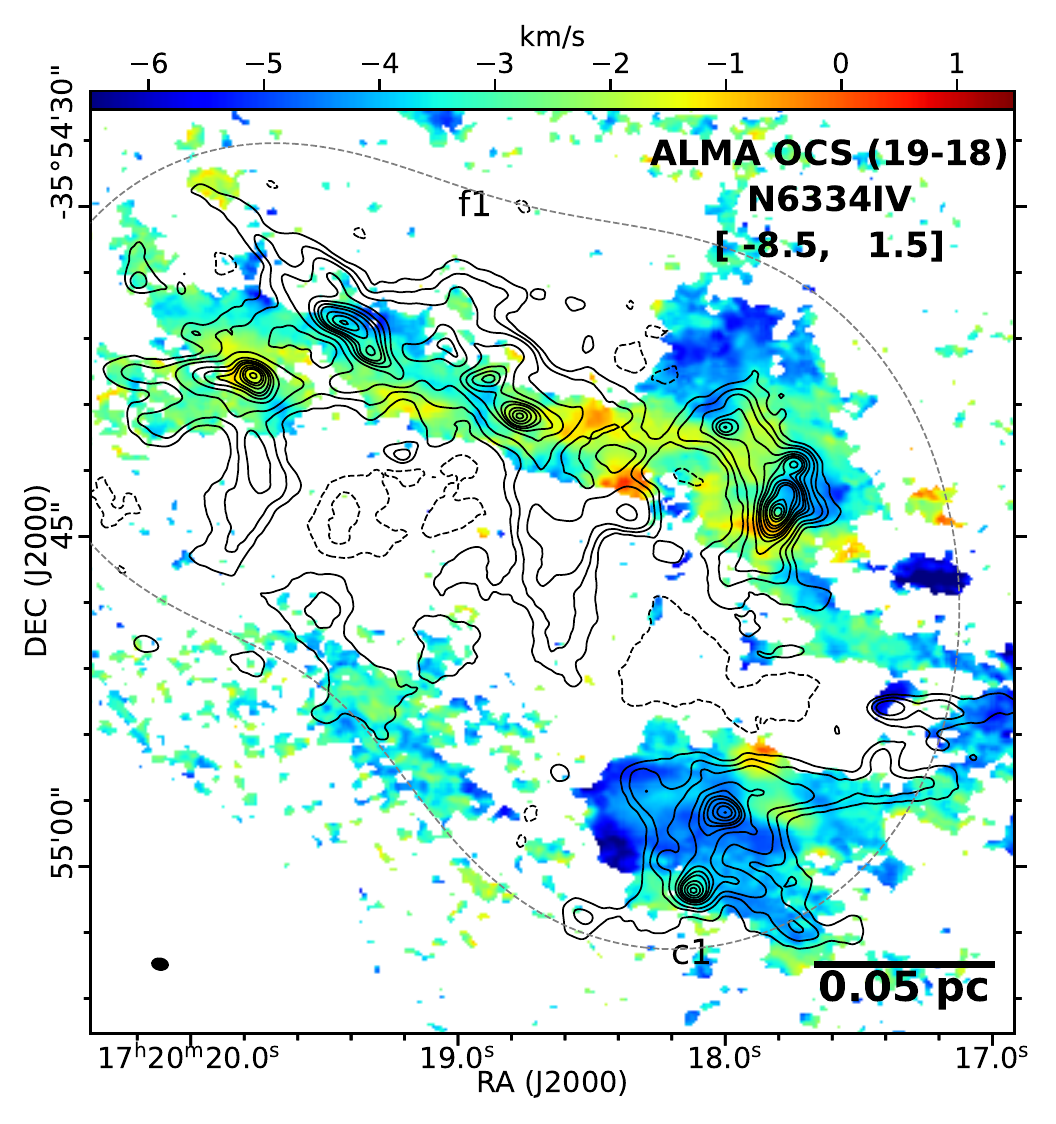}{0.33\textwidth}{(c)}
 }
 \gridline{\fig{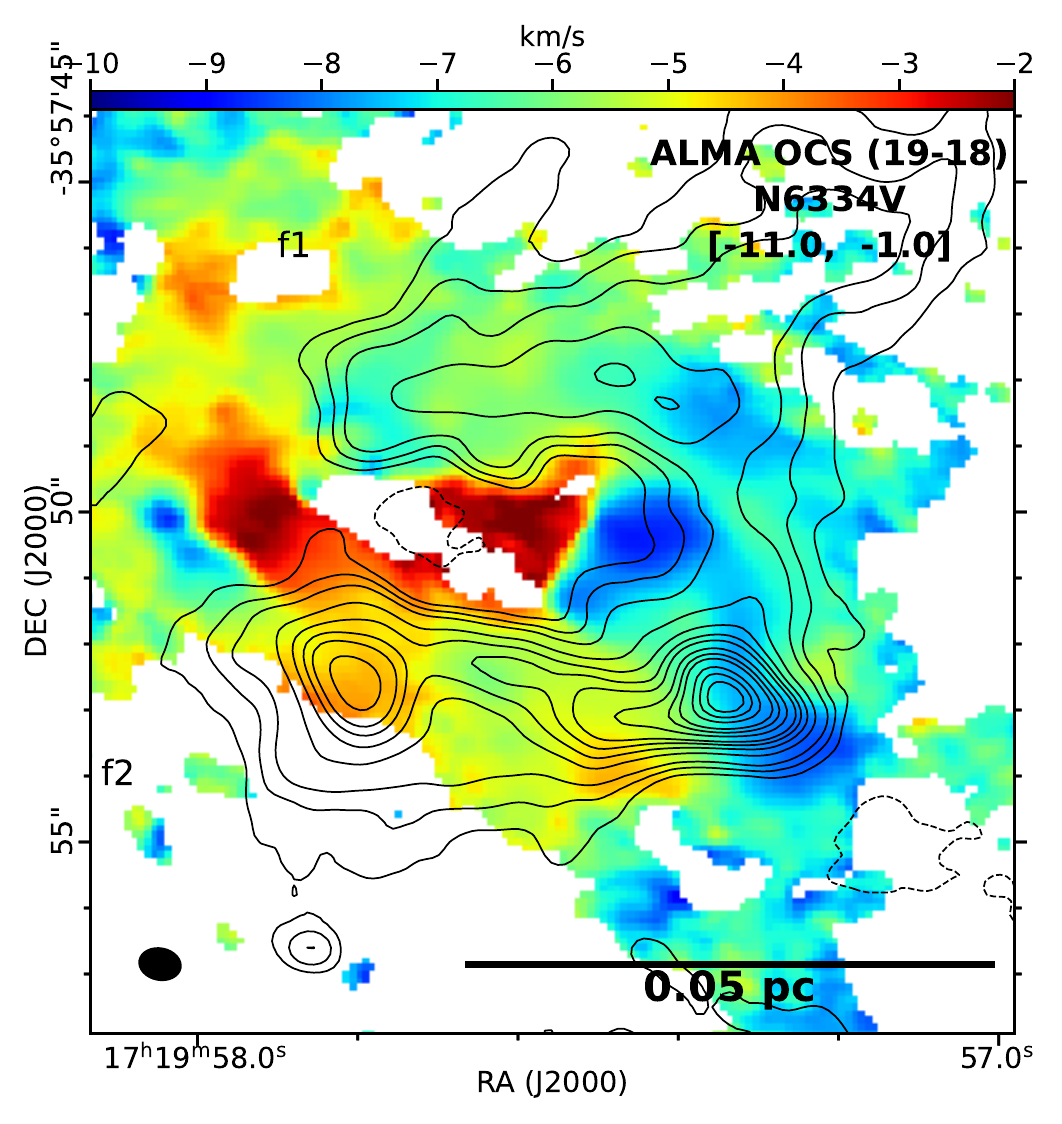}{0.33\textwidth}{(d)}
 \fig{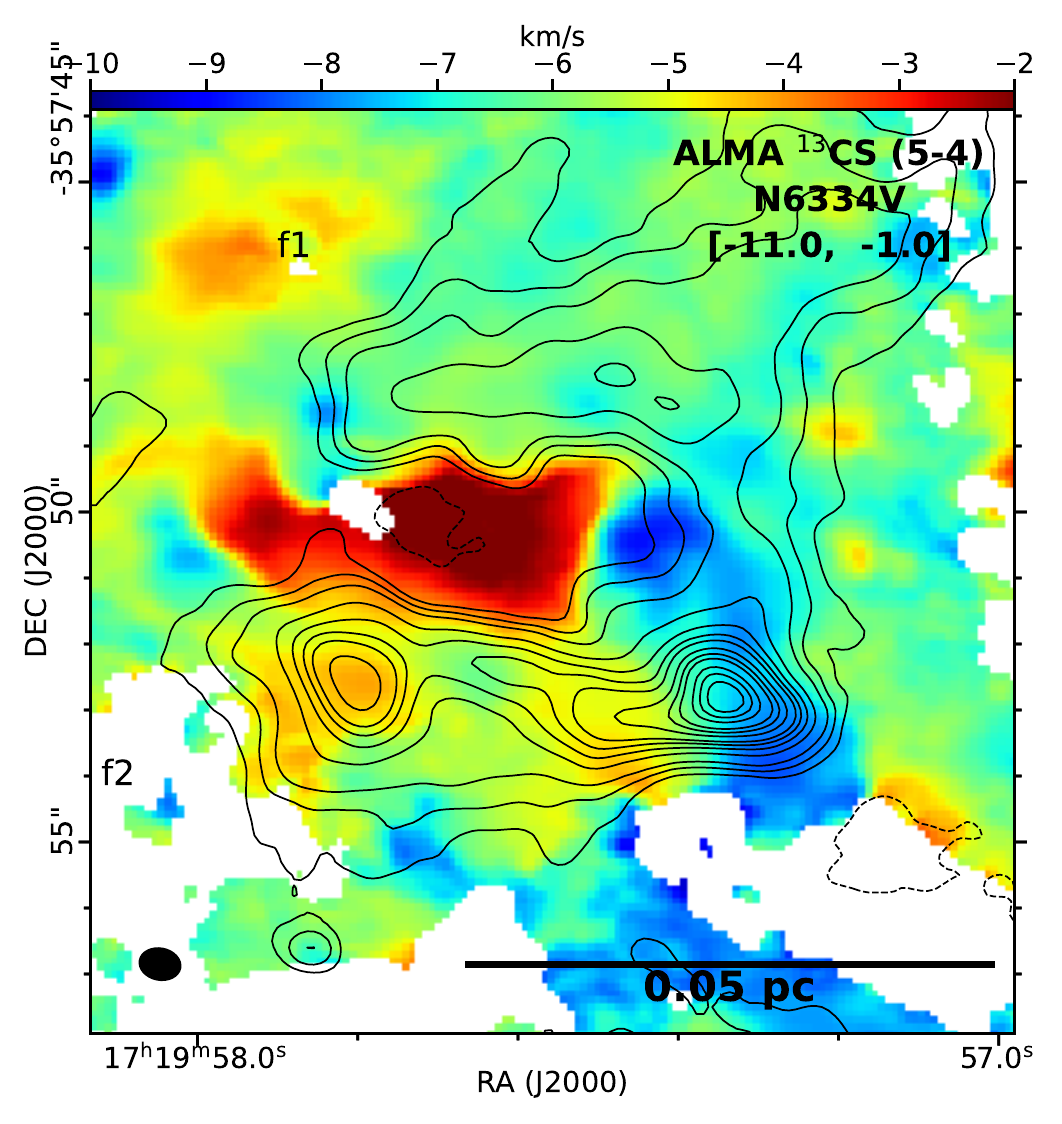}{0.33\textwidth}{(e)}
 }
\gridline{\fig{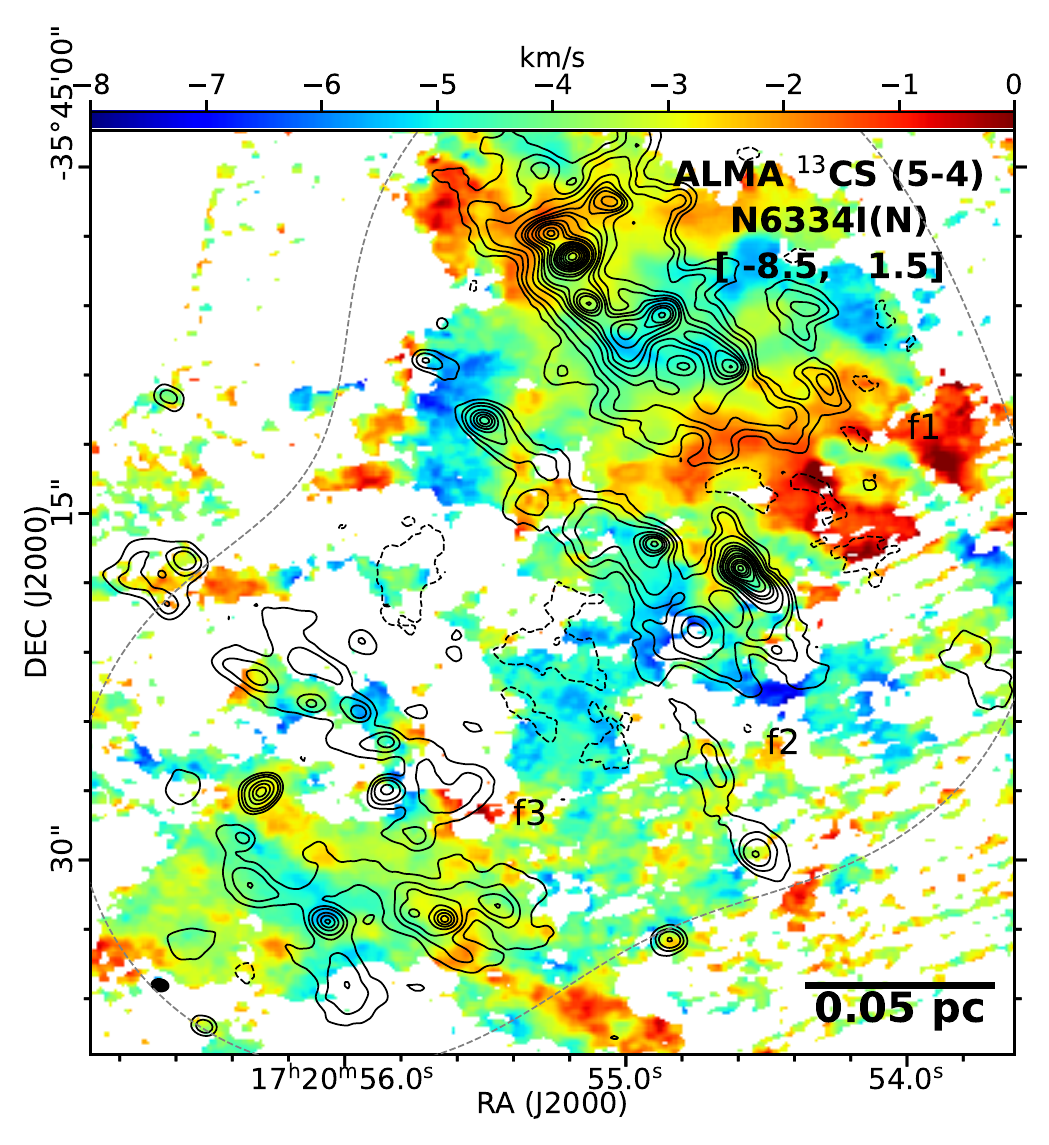}{0.33\textwidth}{(f)}
 \fig{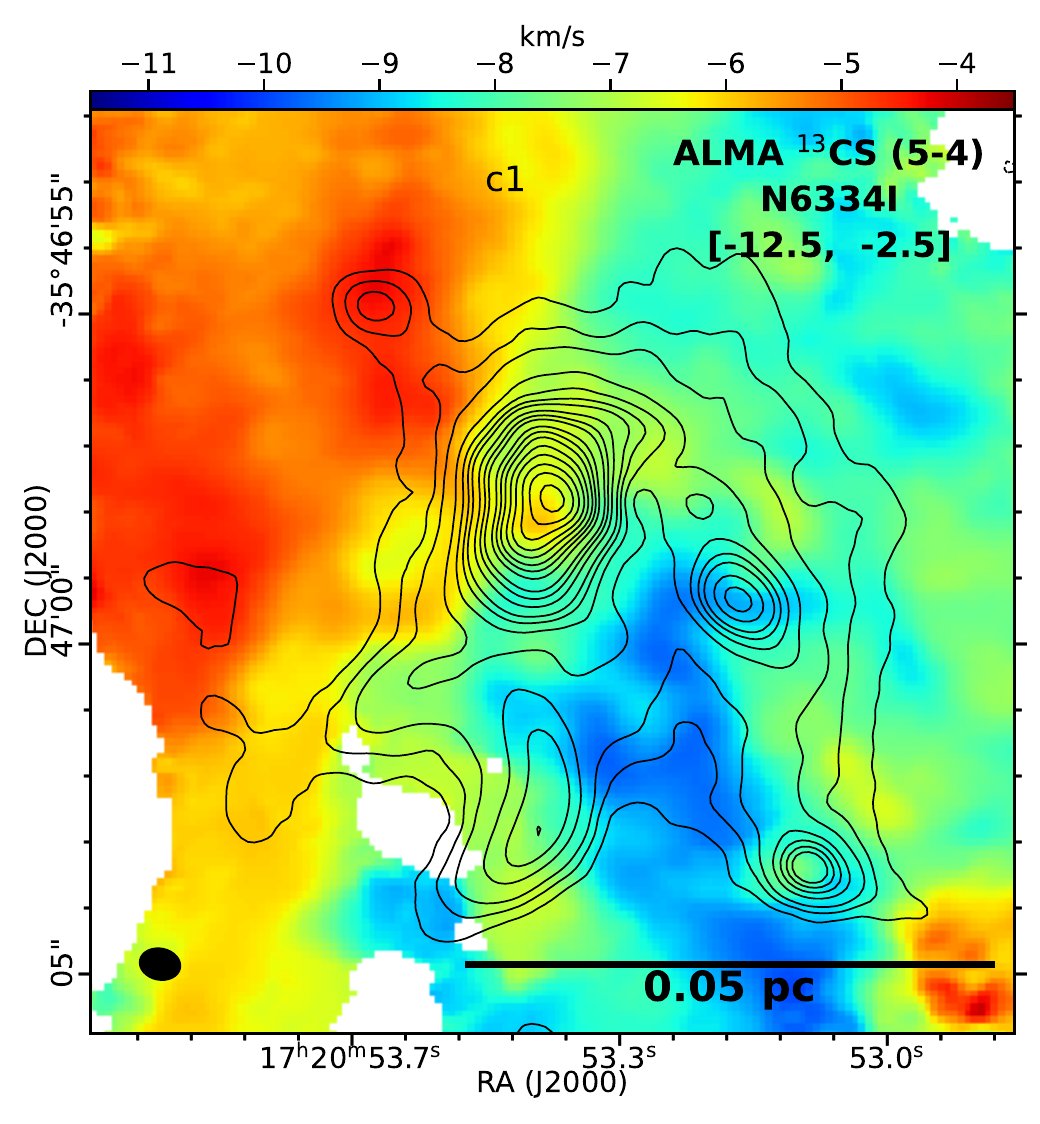}{0.33\textwidth}{(g)}
 \fig{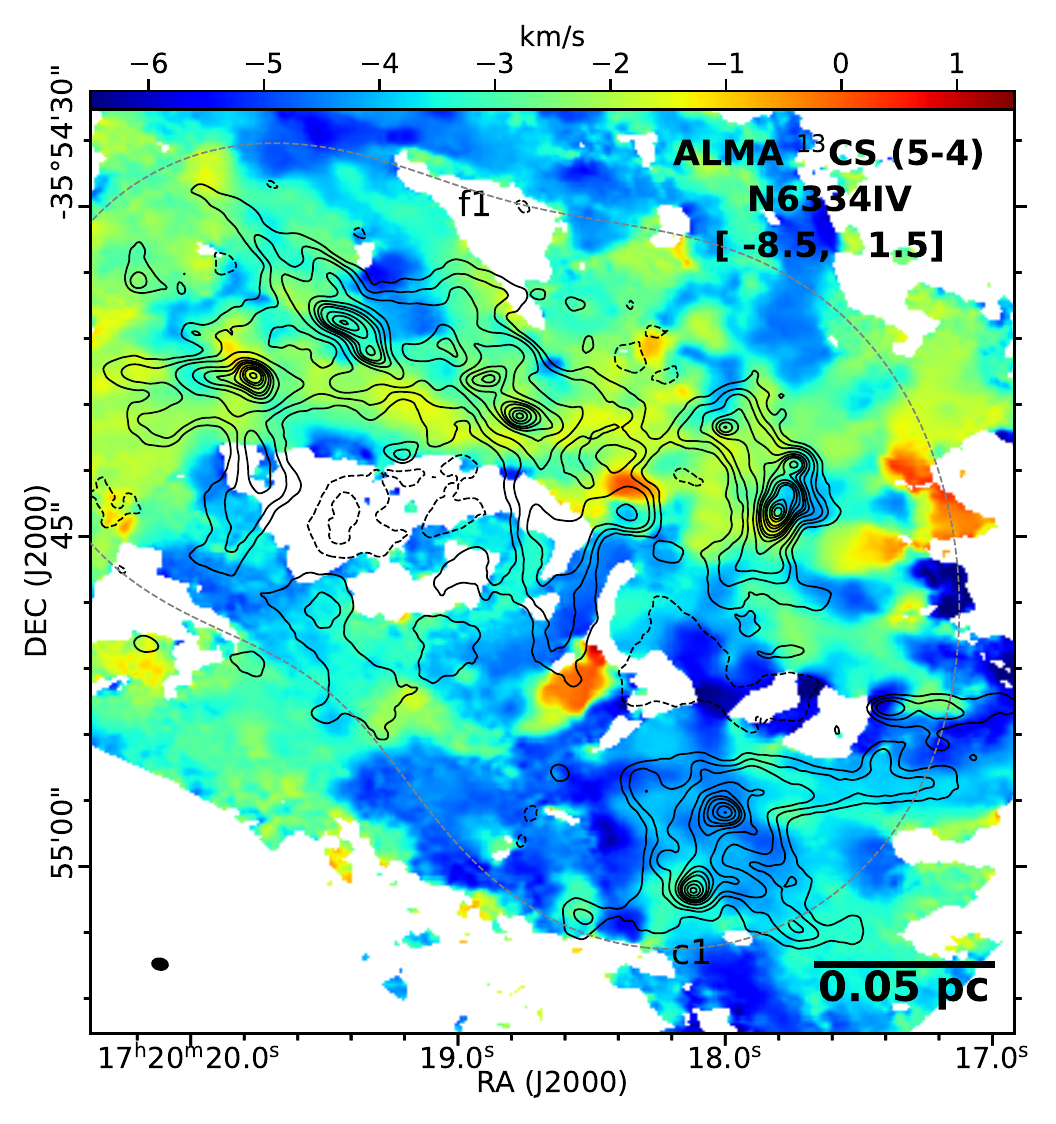}{0.33\textwidth}{(h)}
 }
\caption{Velocity centroid maps of ALMA OCS ((a)-(d)) and $^{13}$CS ((e)-(h)) observations. The contour levels correspond to the ALMA 1.3 mm dust continuum map \citep{2023ApJ...945..160L}. Contour levels are ($\pm$3, 6, 10, 20, 30, 40, 50, 70, 90, 110, 150, 180, 210, 250, 290, 340, 390, 450) $\times \sigma_{I}$, where $\sigma_{I}=$ 1.4, 5.2, 1.2, and 1.0 mJy beam$^{-1}$ for N6334I(N), I, IV, and V, respectively. The synthesized beam and a scale bar are indicated in the lower left and lower right corners of each panel, respectively. Panels were previously shown in \citet{2023ApJ...945..160L} and are reproduced with permission.  \label{fig:N6334_alma_line_m1}}
\end{figure*}

Figures \ref{fig:N6334_large_line_m1}, \ref{fig:N6334_alma_S_line_m1}, \ref{fig:N6334_sma_line_m1}, and \ref{fig:N6334_alma_line_m1} present the intensity-weighted velocity centroid (i.e., moment 1) maps of the molecular lines used in our analysis. 

\bibliography{N6334vdf}{}
\bibliographystyle{aasjournal}


\end{CJK*}
\end{document}